\documentclass[aps,prb,showpacs,twocolumn,floatfix]{revtex4}
\usepackage{graphicx}
\usepackage{amsmath}
\usepackage{amsfonts}
\usepackage{amssymb}
\usepackage{bm}
\usepackage{dcolumn}

\begin{document}

\title{First-principles scattering matrices for spin-transport}
\author{K. Xia}
\altaffiliation[Present address:]{State Key Laboratory for Surface
Physics, Institute of Physics, Chinese Academy of Sciences, P. O.
Box 603, Beijing 100080, China.}
\author{M. Zwierzycki}
\altaffiliation[Present address:]
{Max-Planck-Institut f\"{u}r Festk\"{o}rperforschung, Heisenbergstr. 1,
D-70569 Stuttgart, Germany.}
\altaffiliation[Permanent address:]{Institute of
Molecular Physics, P.A.N., Smoluchowskiego 17, 60-179 Pozna\'n,
Poland.}
\author{M. Talanana}
\author{P. J. Kelly}
\affiliation{Faculty of Science and Technology and MESA$^{+}$
Research Institute, University of Twente, P.O. Box 217, 7500 AE
Enschede, The Netherlands}
\author{G. E. W. Bauer}
\affiliation{Kavli Institute of NanoScience, Delft University of
Technology, Lorentzweg 1, 2628 CJ Delft, The Netherlands }

\date{\today }

\begin{abstract}
Details are presented of an efficient formalism for calculating
transmission and reflection matrices from first principles in layered
materials. Within the framework of spin density functional theory and
using tight-binding muffin-tin orbitals, scattering matrices are
determined by matching the wave-functions at the boundaries between
leads which support well-defined scattering states and the scattering
region. The calculation scales linearly with the number of principal
layers $N$ in the scattering region and as the cube of the number of atoms
$H$ in the lateral supercell. For metallic systems for which the required
Brillouin zone sampling decreases as $H$ increases, the final scaling goes
as $H^{2}N$. In practice, the efficient basis set allows scattering
regions for which $H^{2}N \sim 10^{6}$ to be handled.
The method is illustrated for Co/Cu multilayers and single interfaces
using large lateral supercells (up to $20\times 20$) to model interface
disorder. Because the scattering states are explicitly found,
``channel decomposition'' of the interface scattering for clean and
disordered interfaces can be performed.
\end{abstract}

        
\maketitle


\section{INTRODUCTION}
\label{sec:intro}

One of the most important driving forces in condensed matter physics
in the last thirty years has been the controlled growth of layered
structures so thin that interface effects dominate bulk properties and
quantum size effects can be observed. In doped semiconductors, the
large Fermi wavelength of mobile charge carriers made it possible to
observe finite size effects for layer thicknesses on a micron scale.
Much thinner layers must be used in order to make such observations in
metals because Fermi wavelengths are typically of the order of an
interatomic spacing. Nevertheless, following rapidly on the heels of a
number of important discoveries in semiconductor heterostructures,
interface-dominated effects such as interface magnetic anisotropy,
oscillatory exchange coupling and giant magnetoresistance (GMR) were
found in artificially layered transition metal materials. Reflecting
the shorter Fermi wavelength, the characteristic length scale is of
order of nanometers.

Our main purpose in this paper is to give details of a scheme we have
developed which is suitable for studying mesoscopic transport in
inhomogeneous, mainly layered, transition metal magnetic materials.
In the context of a large number of schemes designed to study transport
either from first-principles\cite{
Schep:prl95,
Zahn:prl95,
Weinberger:jp96,
Schep:prb97,
Zahn:prl98,
vanHoof:prb99,
MacLaren:prb99,
Kudrnovsky:prb00,
Xia:prb01,
Riedel:prb01,
Taylor:prb01,
Brandbyge:prb02,
Drchal:prb02,
Wortmann:prb02a,Wortmann:prb02b,
Weinberger:prp03,
Thygesen:prb03,
Mavropoulos:prb04}
or based upon electronic structures calculated from first-principles\cite{
Mathon:prb97a,
Mathon:prb97b,
Tsymbal:jp97,
Sanvito:prb99,
Velev:prb03,Velev:prb04}
we will require our computational scheme to be (i)
physically transparent, 
(ii) first-principles, requiring no free parameters, 
(iii) capable of handling complex electronic structures characteristic of
transition metal elements and (iv) very efficient in order to be able to
handle lateral supercells to study layered systems with different lattice
parameters and to model disorder very flexibly. 
A tight-binding (TB) muffin-tin-orbital (MTO) implementation of the
Landauer-B\"{u}ttiker formulation of transport theory within the
local-spin-density approximation (LSDA) of density-functional-theory (DFT)
will satisfy these requirements.

\begin{figure}[tbp]
\includegraphics[scale=0.40,angle=0,clip=true]{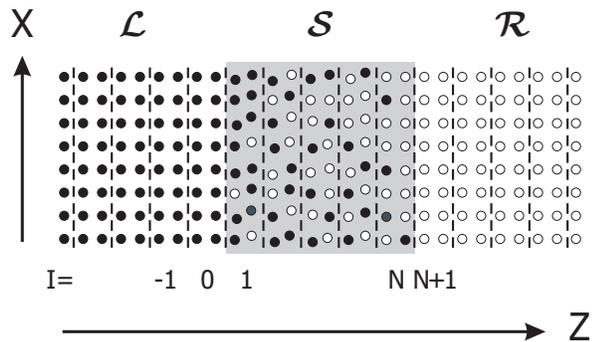}
\caption{Sketch of the configuration used in the Landauer-B\"{u}ttiker
transport formulation to calculate the two terminal conductance.
A (shaded) scattering region ($\mathcal{S}$) is sandwiched by
left- ($\mathcal{L}$) and right-hand ($\mathcal{R}$) leads which
have translational symmetry and are partitioned into principal layers
perpendicular to the transport direction. The scattering region
contains $N$ principal layers but the structure and chemical
composition are in principle arbitrary.}
\label{fig:LSR}
\end{figure}

Because wave transport through interfaces is naturally described
in terms of transmission and reflection, the Landauer-B\"{u}ttiker
(LB) transmission matrix formulation of electron transport gained
rapid acceptance as a powerful tool in the field of mesoscopic
physics,\cite{Imry:02,Datta:95} once the controversies surrounding
the circumstances under which different expressions should be used
had been resolved. \cite{Imry:02} The two-terminal conductance of
a piece of material is measured by attaching leads on either side,
passing a current through these leads and measuring the potential
drop across the scattering region. In the LB formulation of
transport theory, the conductance $G$ is expressed in terms of a
transmission matrix $t\equiv t(E_{F})$
\begin{equation}
G=\frac{e^{2}}{h}Tr\{tt^{\dag }\}  \label{eq:LB}
\end{equation}%
where the element $t_{\mu \nu }$ is the probability amplitude that a state
$\left\vert \nu \right\rangle $ in the left-hand lead incident on the
scattering region from the left (see Fig.~\ref{fig:LSR}) is scattered into a
state $\left\vert \mu \right\rangle $ in the right-hand lead.
The trace simply sums over all incident and transmitted ``channels''
$\nu $ and $\mu $ and $\frac{e^{2}}{h}$ is the fundamental unit of
conductance. In much current work on first-principles transport the
conductance is calculated directly from Green's functions expressed in
some convenient localized orbital representation.\cite{Fisher:prb81}
Explicit calculation of the scattering states is avoided by making use
of the invariance properties of a trace. Because we want to make contact
with a large body of theoretical literature\cite{Beenakker:rmp97}
on mesoscopic physics and address a wider range of problems in the
field of spin-dependent transport, we will calculate the
microscopic transmission and reflection matrices $t$ and $r$. By
using a real energy, we will avoid the problems encountered in
distinguishing propagating and evanescent states when a small but
finite imaginary part of the energy is used. The
Landauer-B\"{u}ttiker formalism satisfies our first requirement of
physical transparency.

In developing a scheme for studying transport in transition metal
multilayers, a fundamental difference between semiconductors and
transition metals must be recognized. Transition metal atoms have
two types of electrons with different orbital character. The
\textit{s} electrons are spatially quite extended and, in solids,
form broad bands with low effective masses; they conduct easily.
The \textit{d} electrons are much more localized in space, form
narrow bands with large effective masses and are responsible for
the magnetism of transition metal elements. The ``magnetic''
electrons, however, being itinerant do contribute to electrical
transport. The appropriate framework for describing metallic
magnetism, even for the late 3$d$ transition metal elements, is
band theory.\cite{Kubler:00} An extremely successful framework
exists for treating itinerant electron systems from
first-principles and this is the Local Density Approximation (LDA)
of Density Functional Theory (DFT). For band magnetism, the
appropriate extension to spin-polarized systems, the local
spin-density approximation (LSDA) satisfies our second requirement
of requiring no free parameters.\cite{fn:total_en}

Oscillatory exchange coupling in layered magnetic structures was discussed
by Bruno in terms of generalized reflection and transmission matrices \cite%
{Bruno:jmmm93} which were calculated by Stiles
\cite{Stiles:jap96,Stiles:prb96,Stiles:prb00} for realistic electronic structures
using a scheme \cite{Stiles:prb88,Stiles:prl91} based on
linearized augmented plane waves (LAPWs). At an interface between a
non-magnetic and a magnetic metal, the different electronic structures
of the majority and minority spin electrons in the magnetic material
give rise to strongly spin-dependent reflection.
\cite{Schep:prl95,Schep:prb98} Schep used transmission and reflection
matrices calculated from first-principles with an embedding surface
Green's function method \cite{vanHoof:97} to calculate spin-dependent
interface resistances for specular Co/Cu interfaces embedded in
diffusive bulk material.\cite{Schep:prb97} The resulting good
agreement with experiment indicated that interface disorder is less
important than the spin-dependent reflection and transmission from a
perfect interface.  Calculations of domain wall resistances as a
function of the domain wall thickness illustrated the usefulness of
calculating the full scattering matrix.
\cite{vanHoof:jmmm98,vanHoof:prb99} However, the LAPW basis set used
by Stiles and Schep was computationally too expensive to allow
repeated lateral supercells to be used to model interfaces between
materials with very different, incommensurate lattice parameters or to
model disorder. This is true of all plane-wave based basis sets which
typically require of order 100 plane waves per atom in order to
describe transition metal atom electronic structures reasonably well.

Muffin-tin orbitals (MTO) form a flexible, minimal basis set leading to
highly efficient computational schemes for solving the Kohn-Sham equations
of DFT.\cite{Andersen:prl84,Andersen:85,Andersen:prb86,Andersen:87}
For the close packed structures adopted by the magnetic materials Fe, Co, Ni
and their alloys, a basis set of 9 functions ({\em s, p}, and {\em d} orbitals)
per atom in combination with the atomic
sphere approximation (ASA) for the potential leads to errors in describing
the electronic structure which are comparable to the absolute errors
incurred by using the local density approximation. This should be compared
to typically 100 basis functions per atom required by the more accurate LAPW
method. MTOs thus satisfy our third and fourth requirements of being able to
treat complex electronic structures efficiently.

The tight-binding linearized muffin tin orbital (TB-LMTO) surface Green's
function (SGF) method has been developed to study the electronic structure
of interfaces and other layered systems. When combined with the
coherent-potential approximation (CPA), it allows the electronic structure,
charge and spin densities of layered materials with substitutional disorder
to be calculated self-consistently very efficiently.\cite{Turek:97} In this
paper we describe how we have combined a method for calculating transmission
and reflection matrices based on wave-function matching (WFM), in a form
given by Ando\cite{Ando:prb91} for an empirical tight-binding Hamiltonian,
with a first-principles TB-MTO basis. \cite{Andersen:prb86} Applications of
the method to a number of problems of current interest in spin-transport
have already been given in a number of short publications: to the
calculation of spin-dependent interface resistances where interface disorder
was modelled by means of large lateral supercells; \cite{Xia:prb01} to the
first principles calculation of the so-called mixing conductance parameter
entering theories of current-induced magnetization reversal\cite{Xia:prb02}
and Gilbert-damping enhancement via spin-pumping;\cite{Zwierzycki:prb05}
to a generalized scattering formulation of the suppression of Andreev
scattering at a ferromagnetic/superconducting interface; \cite{Xia:prl02} to
the problem of how spin-dependent interface resistances influence spin
injection from a metallic ferromagnet into a III--V semiconductor.\cite%
{Zwierzycki:prb03,Bauer:prl04} These examples amply demonstrate that the
fourth requirement is well satisfied.

In Sec.~\ref{sec:theory}, we give technical details of the formalism and
illustrate it in Sec.~\ref{sec:Calc} where we calculate the transmission
matrices for clean and disordered Co/Cu interfaces, document a number of
convergence and accuracy issues and give a detailed ``channel-decomposition''
analysis of the scattering in the presence of disorder. A comparison with
other methods is made in Sec.~\ref{sec:discussion}.

\section{THEORY}
\label{sec:theory}

Central to the wave-function matching method for calculating the
transmission and reflection matrices is the equation of motion (EoM) for
electrons with energy $\varepsilon $, relating the vectors of coefficients $%
\mathbf{C}_{I}$ for layers $I-1$, $I$, and $I+1$:
\begin{equation}
{\mathcal{H}}_{I,I-1}\mathbf{C}_{I-1}+({\mathcal{H}}-\varepsilon )_{I,I}%
\mathbf{C}_{I}+{\mathcal{H}}_{I,I+1}\mathbf{C}_{I+1}=0.  \label{eq:EoM}
\end{equation}%
Here, $\mathbf{C}_{I} \equiv C_{Ii}$ describes the wavefunction
amplitude in terms of some localized orbital basis $\left\vert
i\right\rangle $ of dimension $M$ where $i$ labels the atomic
orbital and atom site. [For the muffin-tin orbitals to be outlined
in Sec.~\ref{ssec:MTOs}, $i$ will be a combined index
$\mathbf{R}lm$, where $l$ and\ $m$ are the azimuthal and magnetic
quantum numbers, respectively, of the MTO defined for an
atomic-spheres-approximation (ASA) potential on the site
$\mathbf{R}$.] The EoM does not restrict us to only considering
nearest neighbour interactions since atoms can always be grouped
into layers defined as to be so thick that the interactions
between layers $I$ and $I\pm 2$ are negligible (see
Fig.~\ref{fig:LSR}). Such layers are called \textit{principal
layers}. Their thickness depends on the range of the interactions
which in turn partly depends on the spatial extent of the orbital
basis. It will be minimized by using the highly localized
tight-binding MTO representation.

Consider the situation sketched in Fig.~\ref{fig:LSR} where the
scattering region $\mathcal{S}$ is contacted with left ($\mathcal{L)}$
and right ($\mathcal{R)}$ leads which have perfect lattice periodicity
and support well-defined scattering states. We assume that the ground
state charge and spin densities and the corresponding Kohn-Sham
independent electron potential have already been calculated
self-consistently. The calculation of the scattering matrix can now be
split into two distinct parts. In the first stage, to be discussed in
Sec.~\ref{ssec:Leads}, the eigenmodes of the leads
$\mathbf{u}_{\mu } (= \mathbf{C}_{0}$ for the $\mu$-th mode),
of which there are
$2M$, are calculated using an EoM appropriate to MTOs and making use
of the lattice periodicity. By calculating their \textbf{k }vectors
(which are in general complex)\ and velocities $\upsilon_{\mathbf{k}}$,
 the eigenstates can be classified as being either left-going
$\mathbf{u}_{\mu }(-)$ or right-going $\mathbf{u}_{\mu }(+)$. They form a
basis in which to expand any left- and right-going waves and have the
convenient property that their transformation under a lattice translation in
the leads is easily calculated using Bloch's theorem (with \textbf{k }%
complex). We use the small Roman letters \emph{i,j} to label the
non-orthogonal basis and the small Greek letters $\mu ,\nu $ to label the
lead eigenmodes.

In the second stage discussed in Sec.~\ref{ssec:scatt}, a scattering
region $\mathcal{S}$ is introduced in the layers $1 \leq I \leq {\rm N}$
which mixes left- and right-going lead eigenmodes. The scattering region
can be a single interface, a complex multilayer or a tunnelling junction,
and the scattering can be introduced by disorder or simply by
discontinuities in the electronic structure at interfaces.
The $\nu \rightarrow \mu $ element of the reflection matrix,
$r_{\mu \nu },$ is defined in terms of the ratio of the amplitudes of
left-going and right-going solutions in the left lead (in layer 0 for example)
projected onto the $\nu ^{\text{th}}$ right-going and $\mu ^{\text{th}}$
left-going propagating states (\textbf{k }vector real) renormalized with the
velocities so as to have unit flux. The scattering problem is solved by
direct numerical inversion of a matrix with the leads included as a boundary
condition so as to make finite the matrix which has to be inverted.

\subsection{Muffin Tin Orbitals and the KKR equation}
\label{ssec:MTOs}

Muffin-tin orbitals
\cite{Andersen:prl84,Andersen:85,Andersen:prb86,Andersen:87} (MTO) are
defined for potentials in which space is
divided into non-overlapping atom-centred ``muffin-tin'' spheres inside
which the potential is spherically symmetric and the remaining
``interstitial'' region where the potential is taken to be
constant. The atomic spheres approximation (ASA) is obtained (i) by taking
the kinetic energy in the interstitial region to be zero and (ii) by
expanding the muffin-tin spheres so that they fill all space whereby the
volume of the interstitial region vanishes; for monoatomic solids such
spheres are called atomic Wigner-Seitz (WS) spheres. Inside a WS (or MT)
sphere at $\mathbf{R}$, the solution of the radial Schr\"{o}dinger equation
regular at $\mathbf{R}$, $\phi _{Rl}(\varepsilon ,r_{R})$ can be determined
numerically for energy $\varepsilon $ and angular momentumum $l$ resulting
in the partial wave
\begin{equation}
\phi _{Rlm}(\varepsilon ,\mathbf{r}_{R})\equiv \phi _{RL}(\varepsilon ,%
\mathbf{r}_{R})\equiv \phi _{Rl}(\varepsilon ,r_{R})Y_{lm}(\mathbf{\hat{r}}%
_{R})  \label{eq:PWdef}
\end{equation}%
where $\mathbf{r}_{R}\equiv \mathbf{r-R}$ and \ $r_{R}\equiv \left\vert
\mathbf{r-R}\right\vert $. A continuous and differentiable orbital is
constructed by attaching \ to the partial wave at the sphere boundary
$r_{R}\equiv s_{R}$ a ``tail'' consisting of an appropriate linear
combination of the solutions of the Laplace equation,
\begin{equation}
J_{RL}^{0}(\mathbf{r}_{R})\equiv \;(r_{R}/\omega )^{l}[2(2l+1)]^{-1}Y_{L}(%
\mathbf{\hat{r}}_{R})  \label{eq:J0}
\end{equation}%
and
\begin{equation}
K_{RL}^{0}(\mathbf{r}_{R})\equiv (r_{R}/\omega )^{-l-1}Y_{L}(\mathbf{\hat{r}}%
_{R}),  \label{eq:K0}
\end{equation}%
which are respectively, regular at $\mathbf{R}$ and at infinity. $\omega $
is the average WS radius if the structure contains different atoms. In terms
of the logarithmic derivative of $\phi _{l}(\varepsilon ,r)$ at $r\equiv s$
\begin{equation}
D_{l}(\varepsilon ,s)\equiv \frac{s\phi _{l}^{\prime }(\varepsilon ,s)}{\phi
_{l}(\varepsilon ,s)}  \label{eq:LD}
\end{equation}%
($\phi _{l}^{\prime }(\varepsilon ,s)$ is the radial derivative), the radial
solutions are matched if for $r \geqslant s$,
\begin{align}
\phi _{l}(\varepsilon ,r)& =\frac{l-D_{l}}{2l+1}\left( \frac{s}{\omega }%
\right) ^{l+1}\phi _{l}(\varepsilon ,s)  \notag \\
& \times \left[ K_{l}^{0}(r)-2(2l+1)\left( \frac{\omega }{s}\right) ^{2l+1}(%
\frac{D_{l}+l+1}{D_{l}-l})J_{l}^{0}(r)\right]
\end{align}%
where we drop the explicit $\mathbf{R}$-dependence when this does not give
rise to ambiguity, or in terms of the potential function
\begin{equation}
P_{l}^{0}(\varepsilon )=2(2l+1)
\left( \frac{\omega}{s}\right) ^{2l+1}
\frac{D_{l}(\varepsilon )+l+1}{D_{l}(\varepsilon )-l}
\end{equation}%
and normalization $N_{l}^{0}(\varepsilon )=
\frac{2l+1}{l-D_{l}}
\left( \frac{\omega }{s}\right) ^{l+1} \frac{1}{\phi _{l}(\varepsilon ,s)}$
\begin{equation}
N_{l}^{0}(\varepsilon )\phi _{l}(\varepsilon
,r)=K_{l}^{0}(r)-P_{l}^{0}(\varepsilon )J_{l}^{0}(r)
\end{equation}
By subtracting from the partial wave, both inside and outside the MT sphere,
the $J_{RL}^{0}(\mathbf{r}_{R})$ component which is irregular at infinity, a
function is formed which is continuous, differentiable and regular in all
space, an energy-dependent muffin tin orbital $\chi _{RL}^{0}(\varepsilon ,%
\mathbf{r}_{R})$:
\begin{align}
\chi _{RL}^{0}(\varepsilon ,\mathbf{r}_{R})& =N_{Rl}^{0}(\varepsilon )\phi
_{Rl}(\varepsilon ,\mathbf{r})+P_{Rl}^{0}(\varepsilon )J_{RL}^{0}(\mathbf{r}%
_{R}) & r_{R}& \leqslant s_{R} \\
& =K_{L}^{0}(\mathbf{r}_{R}) & r_{R}& \eqslantgtr s_{R}  \label{eq:MTOdef}
\end{align}%
The tail $K_{RL}^{0}(\mathbf{r}_{R}\mathbf{)}$\ has the desirable property
that closed forms exist for expanding it around a different site $\mathbf{R}%
^{\prime }$ in terms of the regular solutions $J_{R^{\prime }L^{\prime
}}^{0}(\mathbf{r}_{R^{\prime }}\mathbf{),}$%
\begin{equation}
K_{RL}^{0}(\mathbf{r}_{R}\mathbf{)=-}\sum_{L^{\prime }}
J_{R^{\prime }L^{\prime }}^{0}(\mathbf{r}_{R^{\prime }})
S_{R^{\prime }L^{\prime },RL}^{0}
\label{eq:TailExp}
\end{equation}%
The expansion coefficients $S_{R^{\prime }L^{\prime },RL\text{ }}^{0}$form a
so-called canonical structure constant matrix: they do not depend on the
lattice constant, on the MT (or AS) potentials or on energy. Because of the
augmentation with $J_{RL}^{0}(\mathbf{r}_{R})$,\ the resulting MTO is no
longer a solution of the Schr\"{o}dinger equation (SE) inside its own sphere
$\mathbf{R}$. When, however, a solution of the SE is sought in the form of a
linear combination of MTOs centred on different sites,
\begin{equation}
\Psi (\varepsilon ,\mathbf{r})=\sum_{R,L}\chi _{RL}^{0}(\varepsilon ,\mathbf{%
r}_{R})C_{RL}^{0}  \label{eq:LCMTO}
\end{equation}%
then the partial wave solution is recovered if the augmenting term $%
J_{RL}^{0}(\mathbf{r}_{R})$ on site $\mathbf{R}$ is cancelled by the tails
of MTOs centred on all other sites $\mathbf{R}^{\prime }\neq \mathbf{R}$,
expanded about $\mathbf{R.}$ The condition for this to occur is the
``tail-cancellation'' condition:
\begin{equation}
\sum_{R^{\prime },L^{\prime }}\left[ P_{RL}^{0}(\varepsilon )\delta
_{RR^{\prime }}\delta _{LL^{\prime }}-S_{RL,R^{\prime }L^{\prime }}^{0}\right]
C_{R^{\prime }L^{\prime }}^{0}=0.  \label{eq:PmS}
\end{equation}%
All of the information about the structural geometry of the system
under investigation is contained in the structure constant matrix
$S_{RL,R^{\prime }L^{\prime }}^{0}$ while all of the information about
the atomic species on site $\mathbf{R}$ needed to calculate the
electronic structure (eigenvalues and eigenvectors) is contained in the
potential functions $ P_{RL}^{0}(\varepsilon )$. These are determined
by solving the radial Schr\"{o}dinger equation for the corresponding
spherically symmetrical atomic sphere potential for energy $\varepsilon$
and angular momentum $l$.

A disadvantage of these ``conventional'' MTOs is their infinite range.
However, there is a remarkably simple generalization of the MTOs which allows
their range to be modified by introducing a set of ``screening'' constants
$\alpha _{Rl}$ (not to be confused with the lead eigenmode index) while the
``tail-cancellation'' condition remains essentially unchanged:
\begin{equation}
\sum_{R^{\prime },L^{\prime }}\left[ P_{RL}^{\alpha }(\varepsilon )
\delta_{RR^{\prime }}\delta _{LL^{\prime }}-S_{RL,R^{\prime }L^{\prime }}^{\alpha }%
\right] C_{R^{\prime }L^{\prime }}^{\alpha }=0.  \label{eq:TBPmS}
\end{equation}
$P^{\alpha }(\varepsilon )$ is a diagonal matrix related to
$P^{0}(\varepsilon )$ by
\begin{equation}
P^{\alpha }(\varepsilon )=P^{0}(\varepsilon )+P^{0}(\varepsilon )\alpha
P^{\alpha }(\varepsilon )=\frac{P^{0}(\varepsilon )}{1-\alpha
P^{0}(\varepsilon )},
\end{equation}%
and%
\begin{equation}
S^{\alpha }=S^{0}+S^{0}\alpha S^{\alpha }=S^{0}\left(1-\alpha S^{0}\right)^{-1},
\end{equation}%
For any set of $\alpha _{Rl}$, the energy-dependent MTOs with the
normalization
\begin{equation}
\sum_{R,L}\frac{\omega }{2}\dot{P}_{RL}^{\alpha }(\varepsilon )\left\vert
C_{RL}^{\alpha }\right\vert ^{2}=1,  \label{eq:Norm}
\end{equation}%
form a complete set for the MT (AS) potential used in their construction.
Here, $\dot{P}$ denotes an energy derivative and (\ref{eq:Norm}) follows
from the relation
$N^{\alpha}(\varepsilon)
  =  [(\omega/2) \dot{P}^{\alpha}(\varepsilon)]^{1/2}$.
Sets of parameters $\alpha_{Rl}$ have been found for which the ``screened''
structure constants $S_{RL,R^{\prime }L^{\prime}}^{\alpha }$ have very short
range, decaying exponentially with the interatomic separation.
\cite{Andersen:85} The set of parameters, $\beta _{Rl}$, which yields
the shortest range MTOs is called the ``tight-binding'' (TB) representation.
\cite{Andersen:prl84} For close-packed structures, the range of
$S_{RL,R^{\prime }L^{\prime }}^{\beta }$ is in practice limited to first-
and second-nearest neighbours. This TB set with $\alpha =\beta $ is what
we will use from now, unless stated otherwise, since it will allow us to
define principal layers with a minimal thickness.

For the determination of energy bands $\varepsilon (\mathbf{k}),$ the
tail-cancellation or KKR equations are inconvenient because the
energy-dependence of the potential function makes it necessary to solve (\ref%
{eq:PmS}) or (\ref{eq:TBPmS}) by searching for the roots of a determinant,
which is very time consuming. Much more efficient methods have been
developed based on energy-independent MTOs. However, to study transport we
only need to know $P^{\beta }(\varepsilon )$ for a fixed energy, usually the
Fermi energy. We assume that the Kohn-Sham equations have already been
solved self-consistently (using for example a linearized method) so we have
the potentials from which to calculate the potential functions. Although
(\ref{eq:TBPmS}) can be brought into Hamiltonian form by linearizing the
energy dependent potential function (see Appendix~\ref{sec:velocities}),
we will work directly with the more exact KKR equation.

\subsection{Eigenmodes of the leads}
\label{ssec:Leads}

We will assume that there exists two-dimensional translational symmetry in
the plane perpendicular to the transport direction so that states can be
characterized by a lateral wave vector ${\bf k_{\parallel }}$ in the
corresponding two-dimensional Brillouin zone. The screened KKR equation
\cite{Andersen:85} in the mixed representation of ${\bf k_{\parallel }}$
and real space layer index $I$ (see Fig.~\ref{fig:LSR}) is
\begin{equation}
-S_{I,I-1}^{\bf k_{\parallel }}\mathbf{C}_{I-1}+\left( P_{I,I}(\varepsilon
)-S_{I,I}^{\bf k_{\parallel }}\right) \mathbf{C}_{I}-S_{I,I+1}^{\bf k_{\parallel }}%
\mathbf{C}_{I+1}=0,  \label{eq:PSEoM}
\end{equation}%
where $\mathbf{C}_{I}\equiv C_{Ii}\equiv C_{IRlm}$ is a $(l_{\max }+1)^{2}H$
$\equiv M$ dimensional vector describing the amplitudes of the $I$-th layer
with $H$ sites and $(l_{\max }+1)^{2}$ orbitals per site. $P_{I,I}$
and $S_{I,J}$ are $M\times M$ matrices. $P_{I,I}$ is a diagonal matrix of
potential functions characterizing the AS potentials of layer $I$ and
\begin{equation}
S_{I,J}^{\bf k_{\parallel }}
     = \sum_{ {\bf T} \in  \{ {\bf T}_{I,J}\}}
           S^{\beta }  ({\bf T})
               e^{i {\bf k_{\parallel }.T}},
\end{equation}%
where $\left\{ {\bf T}_{I,J}\right\} $ denotes the set of vectors that
connect one lattice site in the $I$-th layer with all lattice sites in
the $J$-th layer.

By analogy with (\ref{eq:EoM}), equation~(\ref{eq:PSEoM}) is the
equation of motion we will use to calculate the amplitudes of right-
and left-going waves which determine the scattering matrix. We will
solve it for a fixed value of $\varepsilon $ (usually $\varepsilon _{F}$),
and some $\mathbf{k}_{\parallel }$ to find
$k_{\mu}(\varepsilon ,\mathbf{k}_{\parallel })$ the
component of the Bloch vector in the transport direction.
To keep the notation simple, explicit reference to the ${\bf k_{\parallel }}$
and $\varepsilon $ dependence will be omitted from now on.
The formalism to be described in the following can be applied to any
electronic structure code based on the KKR equation~(\ref{eq:PSEoM}),
such as third-generation TB-LMTO.\cite{Andersen:prb00,Tank:pssb00,Andersen:00}

Let us first consider the Bloch states in the ideal lead. To obtain linearly
independent solutions, we set $\mathbf{C}_{I}=\lambda ^{I}\mathbf{C}_{0}$,
since in a periodic potential the wave function should satisfy Bloch's
theorem. The potential function matrix is the same for all unit cells. The
structure constant matrix depends only on the relative positions and,
because that is how they are defined, there is only coupling between
adjacent principal layers so the equation of motion becomes
\begin{equation}
\left(
\begin{array}{cc}
S_{0,1}^{-1}(P-S_{0,0}) & -S_{0,1}^{-1}S_{1,0} \\
1 & 0%
\end{array}%
\right) \left(
\begin{array}{l}
\mathbf{C}_{I} \\
\mathbf{C}_{I-1}%
\end{array}%
\right) =\lambda \left(
\begin{array}{l}
\mathbf{C}_{I} \\
\mathbf{C}_{I-1}%
\end{array}%
\right) ,
\label{eq:leads}
\end{equation}%
The eigenvalue $\lambda _{\mu }$ can be written in the form
$\lambda_{\mu}=exp(i\mathbf{k}_{\mu} \cdot\mathbf{T}_0)$
with $\mathbf{T}_0$ connecting equivalent sites in adjacent
prinicpal layers.
The wave vector ${\bf k}_{\mu }$ can be decomposed into
${\bf k_{\parallel }}$ and a remainder which is in general not real,
${\bf k}_{\mu } =
({\bf k_{\parallel }},{\bf k}_{\mu }-{\bf k_{\parallel }})$.
Equation~(\ref{eq:leads}) has $2M$
eigenvalues and $2M$ eigenvectors, corresponding to $M$ right-going and $M$
left-going waves. By calculating the wavevectors and velocities  (see Eq.~\eqref{eq:velocity} 
and Appendix~\ref{sec:velocities}) of the lead
eigenmodes, the propagating and evanescent states can be identified and
sorted into right-going or left-going modes.

Letting $\mathbf{u}_{1}(-),...,\mathbf{u}_{M}(-)$ denote the left-going
solutions $\mathbf{C}_{0}$ corresponding to eigenvalues $\lambda
_{1}(-),...,\lambda _{M}(-)$ and $\mathbf{u}_{1}(+),...,\mathbf{u}_{M}(+)$
the right-going solutions corresponding to eigenvalues $\lambda
_{1}(+),...,\lambda _{M}(+)$, the matrix $U_{i\mu }(\pm )$ is defined as
\begin{equation}
U(\pm )=(\mathbf{u}_{1}(\pm )...\mathbf{u}_{M}(\pm ))  \label{eq:bigU}
\end{equation}%
and the matrix $\mathbf{\Lambda }(\pm )$ as the diagonal matrix with
elements $\lambda _{\mu }(\pm )$. Following Ando, we next expand any
left- or right-going wave, at $I=0$ for example, as
\begin{equation}
\mathbf{C}_{0}(\pm )=U(\pm )\mathbf{C}(\pm ).  \label{eq:bigC}
\end{equation}%
Note that $\mathbf{C}_{0}$ is a vector whose elements are labelled $i$
while the elements of the vector $\mathbf{C}$ are labelled $\mu $.%
\begin{equation}
F(\pm )\equiv U(\pm )\mathbf{\Lambda }_{\pm }U^{-1}(\pm )  \label{eq:bigF}
\end{equation}
is the matrix of Bloch factors (including evanescent states) transformed
onto the basis $\left\vert i\right\rangle $ and plays a central role in the
following. Knowing it makes it possible to translate a state expressed in
the basis $\left\vert i\right\rangle $ from layer $J$ of the lead to layer
$I$ by
\begin{equation}
\mathbf{C}_{I}(\pm )=F^{I-J}(\pm )\mathbf{C}_{J}(\pm ).  \label{eq:trans}
\end{equation}

\subsection{Scattering problem}
\label{ssec:scatt}

The scattering region $\mathcal{S}$, divided into $N$ principal layers
numbered $1$ to $N$, is now inserted between the left and right leads. The
resulting (scattering region + leads) problem is infinite dimensional in the
real space MTO representation but by making use of their translational
symmetry, the leads can be incorporated as boundary conditions and the
scattering problem can be reduced to a finite problem whose dimension is
determined by the size of the scattering region (number of sites $\times $
number of orbitals per site).

We set about decoupling the scattering region from the leads, first on the
left-hand side, then on the right. The amplitude in the $0$-th layer is
first separated into right- and left- going components $\mathbf{C}_{0}=%
\mathbf{C}_{0}(+)+\mathbf{C}_{0}(-)$. Because there is no additional
scattering in the leads, the right- and left-going components can be
translated to the left by one (principal layer) lattice spacing using the
generalized Bloch factors (\ref{eq:trans}) so the amplitude in layer $-1$
can be related to that in layer $0$ as
\begin{gather}
\mathbf{C}_{-1}=F_{\mathcal{L}}^{-1}(+)\mathbf{C}_{0}(+)+F_{\mathcal{L}%
}^{-1}(-)\mathbf{C}_{0}(-)  \notag \\
=\left[ F_{\mathcal{L}}^{-1}(+)-F_{\mathcal{L}}^{-1}(-)\right] \mathbf{C}%
_{0}(+)+F_{\mathcal{L}}^{-1}(-)\mathbf{C}_{0}.  \label{eq:C-1}
\end{gather}%
allowing us to express $\mathbf{C}_{-1}$ in terms of $\mathbf{C}_{0}$ and $%
\mathbf{C}_{0}(+)$ and so eliminate it from the equation of motion for the $%
0 $-th layer
\begin{equation}
-S_{0,-1}\mathbf{C}_{-1}+\left( P_{0,0}-S_{0,0}\right) \mathbf{C}_{0}-S_{0,1}%
\mathbf{C}_{1}=0,
\end{equation}
which becomes
\begin{eqnarray}
&&(P_{0,0}-\widetilde{S}_{0,0})\mathbf{C}_{0}-S_{0,1}\mathbf{C}_{1}  \notag
\\
&=&S_{0,-1}\left[ F_{\mathcal{L}}^{-1}(+)-F_{\mathcal{L}}^{-1}(-)\right]
\mathbf{C}_{0}(+).
\end{eqnarray}
Here $\mathcal{L}$ denotes the left lead and
$\widetilde{S}_{0,0}=S_{0,0}+S_{0,-1}F_{\mathcal{L}}^{-1}(-)$.
$-S_{0,-1}F_{\mathcal{L}}^{-1}(-)$ is the ``embedding
potential'' for the left lead and the net result is that the equations
of motion have been truncated at layer $0.$

On the right-hand side of the scattering region, we are interested in the
situation where only right-going waves can exist in the $\left( N+1\right) $%
-th layer, so
\begin{equation}
\mathbf{C}_{N+2}=F_{\mathcal{R}}(+)\mathbf{C}_{N+1}(+)  \label{eq:CN+2}
\end{equation}
allowing us to eliminate $\mathbf{C}_{N+2}$\textbf{\ }from the EoM for
$\mathbf{C}_{N+1}$
\begin{equation}
(P_{N+1,N+1}-\widetilde{S}_{N+1,N+1})\mathbf{C}_{N+1}-S_{N+1,N}\mathbf{C}%
_{N}=0,  \label{hamnth}
\end{equation}%
where $\widetilde{S}_{N+1,N+1}=S_{N+1,N+1}+S_{N+1,N+2}F_{\mathcal{R}}(+)$
and $-S_{N+1,N+2}F_{\mathcal{R}}(+)$ is the embedding potential for the
right lead.

\begin{widetext}

Making use of the lead boundary conditions, the tail cancellation condition
for the scattering problem in real space is given by the set of
inhomogeneous linear equations
\begin{eqnarray}
\left( \begin{array}{cccccc}
(P -\widetilde{S})_{0,0} & -S_{0,1}    & 0           & \cdots & 0           & 0 \\
-S_{1,0}                 & (P-S)_{1,1} & -S_{1,2}    & \cdots & 0           & 0 \\
0                        & -S_{2,1}    & (P-S)_{2,2} & \cdots & \vdots      & 0 \\
\vdots                   & \vdots      & \cdots      & \ddots & \vdots      & 0 \\
0                        & 0           & \cdots      & \cdots & (P-S)_{N,N} & -S_{N,N+1} \\
0                        & 0           & 0           & \cdots & -S_{N+1,N}  & (P-\widetilde{S})_{N+1,N+1}
\end{array} \right)
\left( \begin{array}{l}
\mathbf{C}_{0} \\
\mathbf{C}_{1} \\
\mathbf{C}_{2} \\
\vdots \\
\mathbf{C}_{N} \\
\mathbf{C}_{N+1}
\end{array} \right) & & \notag \\
\equiv
\left( \mathbf{P-}\widetilde{\mathbf{S}}\right)
\left(
\begin{array}{l}
\mathbf{C}_{0} \\
\mathbf{C}_{1} \\
\mathbf{C}_{2} \\
\vdots \\
\mathbf{C}_{N} \\
\mathbf{C}_{N+1}
\end{array}
\right)
= \left(
\begin{array}{c}
S_{0,-1}\left[ F_{\mathcal{L}}^{-1}(+)-F_{\mathcal{L}}^{-1}(-)\right] \mathbf{%
C}_{0}(+) \\
0 \\
0 \\
\vdots \\
0 \\
0%
\end{array}
\right) & &  \label{eq:green}
\end{eqnarray}%
which can be solved in terms of
$\mathbf{g=}\left( \mathbf{P-}\widetilde{\mathbf{S}}\right) ^{-1}$
\begin{equation*}
\left(
\begin{array}{l}
\mathbf{C}_{0} \\
\mathbf{C}_{1} \\
\mathbf{C}_{2} \\
\vdots \\
\mathbf{C}_{N} \\
\mathbf{C}_{N+1}%
\end{array}%
\right) =\mathbf{g}\left(
\begin{array}{c}
S_{0,-1}\left[ F_{\mathcal{L}}^{-1}(+)-F_{\mathcal{L}}^{-1}(-)\right] \mathbf{%
C}_{0}(+) \\
0 \\
0 \\
\vdots \\
0 \\
0%
\end{array}%
\right)
\end{equation*}
This treatment is very similar to the widely used surface Green function
method.\cite{Khomyakov:prb05} The boundary conditions in (\ref{eq:green})
are explicitly defined by considering the Bloch wave coming from the
left-hand side while for conventional retarded or advanced Green functions
the boundary conditions are specified by an infinitesimal imaginary part
of the energy parameter $\varepsilon $.

We are now in a position where we can relate the outgoing wave amplitude
in the right electrode to the incoming wave in the left electrode through
the Green function by
\begin{equation}
\mathbf{C}_{N+1}(+) =
\mathbf{C}_{N+1} =
g_{N+1,0}S_{0,-1}
\left[ F_{\mathcal{L}}^{-1}(+)-F_{\mathcal{L}}^{-1}(-)\right]
\mathbf{C}_{0}(+).
\end{equation}%
Using the transformation between the eigenstates and the localized basis
functions $U_{i\alpha }(\pm )$, we obtain the transmission and reflection
matrix elements\cite{Ando:prb91}
\begin{equation}
t_{\mu \nu }=\left( \frac{\upsilon _{\mu }}{\upsilon _{\nu }}\right)^{1/2}
\left\{ U_{\mathcal{R}}^{-1}(+)g_{N+1,0}S_{0,-1}
\left[ F_{\mathcal{L}}^{-1}(+)-F_{\mathcal{L}}^{-1}(-)
\right] U_{\mathcal{L}}(+)\right\}_{\mu \nu }
\label{trale}
\end{equation}
\begin{equation}
r_{\mu \nu }=\left( \frac{\upsilon _{\mu }}{\upsilon _{\nu }}\right)^{1/2}
\left\{ U_{\mathcal{L}}^{-1}(-)
\left\langle g_{0,0}S_{0,-1}
\left[ F_{\mathcal{L}}^{-1}(+)-F_{\mathcal{L}}^{-1}(-) \right] - 1\right\rangle
U_{\mathcal{L}}(+)\right\}_{\mu \nu },
\label{refle}
\end{equation}
where $\mu$ and $\nu $ label Bloch states and $\upsilon_{\mu }$,
$\upsilon_{\mu }$ are the components of the corresponding group
velocities in the transport direction.
Similarly, an incident wave from the right side is transmitted or
reflected as
\begin{equation}
t_{\mu \nu }^{\prime }=
\left( \frac{\upsilon _{\mu }}{\upsilon _{\nu }}\right) ^{1/2}
\left\{ U_{\mathcal{L}}^{-1}(-)g_{0,N+1}S_{N+1,N+2}
\left[ F_{\mathcal{R}}(-)-F_{\mathcal{R}}(+)\right] U_{\mathcal{R}}(-)
\right\}_{\mu\nu }
\label{trari}
\end{equation}
\begin{equation}
r_{\mu \nu }^{\prime }=
\left( \frac{\upsilon _{\mu }}{\upsilon _{\nu }}\right) ^{1/2}
\left\{ U_{\mathcal{R}}^{-1}(+)
\left\langle
g_{N+1,N+1}S_{N+1,N+2}\left[ F_{\mathcal{R}}(-)-F_{\mathcal{R}}(+)\right] -1
\right\rangle
U_{\mathcal{R}}(-) \right\} _{\mu \nu }.
\label{refri}
\end{equation}
The group velocities in (\ref{trale})-(\ref{refri}) are determined using the
expression
\begin{equation}
  \upsilon_{\mu }(\pm)=\frac{id}{\hbar}\left[
    \mathbf{u}^{\dagger}_{\mu}(\pm)S_{I,I+1}^{\mathbf{k}_{\parallel}}
    \mathbf{u}_{\mu}(\pm)\lambda_{\mu}-\mathrm{h.c.}
  \right]
  \label{eq:velocity}
\end{equation}
which is derived in Appendix A. Here, $I$ and $I+1$ denote
neighbouring principal layers in either left or right lead, 
$d= {\bf T}_0 \cdot \hat{n}$ is the distance between equivalent monolayers in
adjacent principal layers and $\hat{n}$ is a unit vector in the
transport direction.
\end{widetext}

\subsection{Disorder}
\label{ssec:Supercells}

Interfaces between materials with different lattice parameters
\cite{Xia:prl02} and disordered interfaces\cite{Xia:prb01,Zwierzycki:prb03}
can be modelled very flexibly using lateral supercells.
This approach allows us to study the effect of various types of disorder
on transport properties, ranging from homogeneous interdiffusion (alloying)
to islands, steps etc.
The supercell description of disorder becomes formally exact in the
limit of infinitely large supercells. In practice, satisfactory
convergence is achieved for supercells of quite moderate size
(see Sec. \ref{ssec:IntDis}).

\subsubsection*{Leads}
The factor limiting the ``size'' of scattering problem which can be handled
in practice is the rank of the 
blocks of the block-tridiagonal
scattering matrix in \eqref{eq:PSEoM}, which is proportional to the number
of atoms in the supercell.
If performed straightforwardly in the manner outlined in Sec.~\ref{ssec:Leads},
the solution of the lead equation \eqref{eq:leads} involves solving a
non-Hermitian eigenvalue problem whose rank is twice as large.
Unless use is made of the greater translational symmetry present in the
leads, this can become the limiting step in the whole calculation.
Doing so makes it possible to reduce the dimension of the lead state
calculation to a size determined by the dimension of a primitive unit
cell which is usually negligible.

\begin{figure}[bp]
\includegraphics[scale=0.60,clip=true]{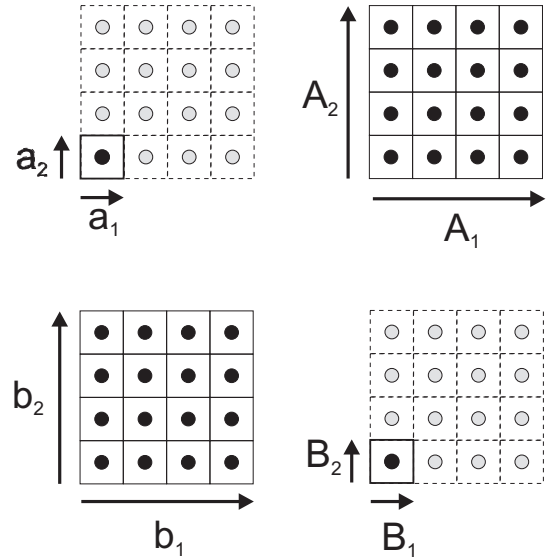}
\caption{Illustration of lateral supercells and corresponding 2D
interface Brillouin zones. Top panel: lattice vectors for a
primitive unit cell containing a single atom (lhs) and a $4 \times
4$ supercell (rhs). Bottom panel: a single k-point in the BZ (rhs)
corresponding to the $4 \times 4$ real-space supercell is
equivalent to $4 \times 4$ k-points in the BZ (lhs) corresponding
to the real-space primitive unit cell.} \label{fig:supercells}
\end{figure}

We consider an $H_1 \times H_2$ lateral supercell defined by the real-space
lattice vectors
\begin{equation}
  \mathbf{A}_1=H_1 \mathbf{a}_1
  \;\;\; \mathrm{and}\;\;\;
  \mathbf{A}_2=H_2 \mathbf{a}_2
  \label{eq:super_bigA}
\end{equation}
where $\mathbf{a}_1$ and $\mathbf{a}_2$ are  the lattice vectors describing
the in-plane periodicity of a primitive unit cell (Fig.~\ref{fig:supercells}).
The cells contained within the supercell are generated by the set of
translations
\begin{align}
  {\mathbf{T}_{\parallel}}_{\mathcal{T}} \in \mathbb{T}=&\left\{
      \mathbf{T}_{\parallel}=  h_1\mathbf{a}_1+ h_2\mathbf{a}_2
    \right. ; \nonumber\\
      & \left. 0 \le h_1 < H_1,\: 0 \le h_2 < H_2
    \right\}
  \label{eq:super_tau}
\end{align}
where $\mathcal{T}=1, \ldots , H_1\times H_2$ is a convenient cell index.
In reciprocal space the supercell Brillouin zone is defined by the
reduced vectors
\begin{equation}
 \mathbf{B}_1= \mathbf{b}_1 / H_1
  \;\;\; \mathrm{and}\;\;\;
  \mathbf{B}_2= \mathbf{b}_2 / H_2
  \label{eq:super_bigB}
\end{equation}
where $\mathbf{b}_1$ and $\mathbf{b}_2$ are the reciprocal lattice
vectors corresponding to the real space primitive unit cell. As a
result the Brillouin zone (BZ) is folded down, as shown schematically
in Fig.~\ref{fig:supercells} (bottom rhs), and the single
$\mathbf{k}^{\mathbb{S}}_{\parallel}$ point ($\mathbb{S}$ is used to
label supercell quantities) in the supercell BZ corresponds to the set
of $H_1 \times H_2$ ${\bf k}_{||}$ points in the original unfolded BZ
\begin{align}
  {\mathbf{k}_{\parallel}}_{\mathcal{K}} \in \mathbb{K}=&\left\{
      \mathbf{k}_{\parallel}= \mathbf{k}^{\mathbb{S}}_{\parallel}+ 
      h_1\mathbf{B}_1+ h_2\mathbf{B}_2
      \right. ; \nonumber\\
      & \left. 0 \le h_1 < H_1,\: 0 \le h_2 < H_2
    \right\}
  \label{eq:super_k}
\end{align}
with $\mathcal{K}=1, \ldots , H_1 \times H_2$.
Solutions associated with different ${\mathbf{k}_{\parallel}}_{\mathcal{K}}$
in the primitive unit cell representaton become different ``bands'' at the
single $\mathbf{k}^{\mathbb{S}}_{\parallel}$ in the supercell representaton.

The indices $\mathcal{T}$ and $\mathcal{K}$ provide a natural means of
describing the supercell-related matrices $U^\mathbb{S}(\pm)$ and
$F^\mathbb{S}(\pm)$ and their inverses in terms of $(H_1\times H_2)^2$
sub-blocks with dimensions defined by the primitive unit cell.
Thus $U^\mathbb{S}_{\mathcal{T} \mathcal{K}}(\pm)$ is the block
containing the amplitudes of the modes associated with
${\mathbf{k}_{\parallel}}_{\mathcal{K}}$ in the
$\mathcal{T}$-th real-space cell.

Solving the single unit cell problem for the set of
$\mathbf{k}_{\parallel}$-points belonging to $\mathbb{K}$
(lhs of Fig.~\ref{fig:supercells}) and using the
Bloch symmetry of the eigenmodes, we get trivially
\begin{equation}
 U^\mathbb{S}_{\mathcal{T} \mathcal{K}}(\mathbf{k}^\mathbb{S}_{\parallel})
    =e^{i {\mathbf{k}_{\parallel}}_{\mathcal{K}}    \cdot
          {\mathbf{T}_{\parallel}}_{\mathcal{T}}           }
    U({\mathbf{k}_{\parallel}}_{\mathcal{K}})
  \label{eq:super_U}
\end{equation}
where $U({\mathbf{k}_{\parallel}}_{\mathcal{K}})$ is the matrix
\eqref{eq:bigU} of modes for a primitive unit cell for
${\mathbf{k}_{\parallel}}_{\mathcal{K}}$ and the $\pm$ qualifier has
been dropped for simplicity. Defining the matrix of phase factors
\begin{equation}
X(\mathbf{k}^\mathbb{S}_{\parallel})=\left(
 \begin{array}{ccc}
   e^{i  {\mathbf{k}_{\parallel}}_1  \cdot {\mathbf{T}_{\parallel}}_1 } & \ldots &
   e^{i  {\mathbf{k}_{\parallel}}_H  \cdot {\mathbf{T}_{\parallel}}_1 }           \\
       \vdots &  & \vdots \\
                          \\
   e^{i  {\mathbf{k}_{\parallel}}_1  \cdot {\mathbf{T}_{\parallel}}_H } & \ldots  &
   e^{i  {\mathbf{k}_{\parallel}}_H  \cdot {\mathbf{T}_{\parallel}}_H }
    \end{array}
    \right)
  \label{eq:super_X}
\end{equation}
with $H \equiv H_1 \times H_2$, and its inverse $Y=X^{-1}$, we can
straightforwardly determine
\begin{equation}
  \left[
    U^{\mathbb{S}}(\mathbf{k}^\mathbb{S}_{\parallel})
  \right]^{-1}_{\mathcal{K} \mathcal{T}} =
  U^{-1}({\mathbf{k}_{\parallel}}_{\mathcal{K}})
                               Y_{\mathcal{K} \mathcal{T}}
  \label{eq:super_Uinv}
\end{equation}
and
\begin{equation}
  F^\mathbb{S}_{\mathcal{T}_1 \mathcal{T}_2}
                    (\mathbf{k}^\mathbb{S}_{\parallel})=
  \sum_{\mathcal{K}}
  X_{\mathcal{T}_1 \mathcal{K}} F({\mathbf{k}_{\parallel}}_{\mathcal{K}})
             Y_{\mathcal{K} \mathcal{T}_2}
  \label{eq:super_F}
\end{equation}

The procedure outlined above for determining the matrices describing the
lead modes scales linearly with the size of the supercell \emph{i.e.}, as
($H_1 \times H_2$) rather than as $(H_1\times H_2)^3$ which is the scaling
typical for matrix operations. Another advantage is that it enables us to
analyse the scattering. By keeping track of the relation between supercell
``bands'' and equivalent eigenmodes at different
${\mathbf{k}_{\parallel}}_{\mathcal{K}}$ (Fig.~\ref{fig:supercells}) we can
straightforwardly obtain from \eqref{trale}-\eqref{refri}
$t_{\mu \nu}   ({\mathbf{k}_{\parallel}}_{\mathcal{K}_1},
                {\mathbf{k}_{\parallel}}_{\mathcal{K}_2})$
and other scattering coefficients. In other words the ``interband''
specular scattering in the supercell picture translates, in the
presence of disorder in the scattering region, into the ``diffuse''
scattering between the $\mathbf{k}_{\parallel}$ vectors belonging
to the $\mathbb{K}$ set.

\section{CALCULATIONS}
\label{sec:Calc}

Even though the theoretical scheme outlined above contains no adjustable
parameters, its practical implementation does involve numerous
approximations, some physical, others numerical, which need to be evaluated.
At present, any workable scheme must be based upon an independent particle
approximation. The results of a transport calculation will be limited by
the extent to which the single particle electronic structures used are
consistent with the corresponding Fermi surfaces determined experimentally
using methods such as de Haas-van Alphen measurements or the occupied and
unoccupied electronic states close to the Fermi energy determined by, for
example, photoelectron spectroscopy.

In this section we examine how various approximations affect our end
results. We begin with the calculation of the scattering states in
bulk Cu and bulk Co (\ref{ssec:CalcsLeads}).
These are then used to study specular scattering from an ideal ordered
Cu/Co(111) interface (\ref{ssec:OrdInt}) after which we describe how we
model disordered interfaces (\ref{ssec:IntDis}) and how the results can
be analysed.

\subsection{Leads}
\label{ssec:CalcsLeads}

For a crystalline conductor with Bloch translational symmetry, each
state at the Fermi energy can move unhindered through the solid so
that the transmission matrix is diagonal with
$\left\vert t_{\mu \nu }\right\vert ^{2}=\delta _{\mu \nu }$.
In this \textit{ballistic} regime, (\ref{eq:LB}) reduces to
\begin{equation}
G^{\sigma }(\hat{n})
       =\frac{e^{2}}{h}
                \sum_{\mu k_{\parallel }}
                     |t_{\mu \mu}^{\sigma }({\bf k}_{\parallel })|^{2}
       =\frac{e^{2}}{h} N^{\sigma }(\hat{n}).
       \label{eq:Sharvin}
\end{equation}
and calculation of the so-called \textit{Sharvin} conductance becomes
a matter of counting the number of modes (channels) propagating in the
transport direction $\hat{n}$, denoted in \eqref{eq:Sharvin} as
$N^{\sigma }(\hat{n})$. To solve \eqref{eq:leads} in practice, the
orbital angular momentum expansions in (\ref{eq:TailExp}) and
(\ref{eq:LCMTO}), which are in-principle infinite, must be truncated
by introducing some cutoff in $l$, denoted $l_{\max }$.  Usually, a
value of $l_{\max }=2$ or $3$ is used, corresponding to \textit{spd}-
or \textit{spdf}-bases.

\begin{figure}[tbp]
\includegraphics[scale=0.35,clip=true]{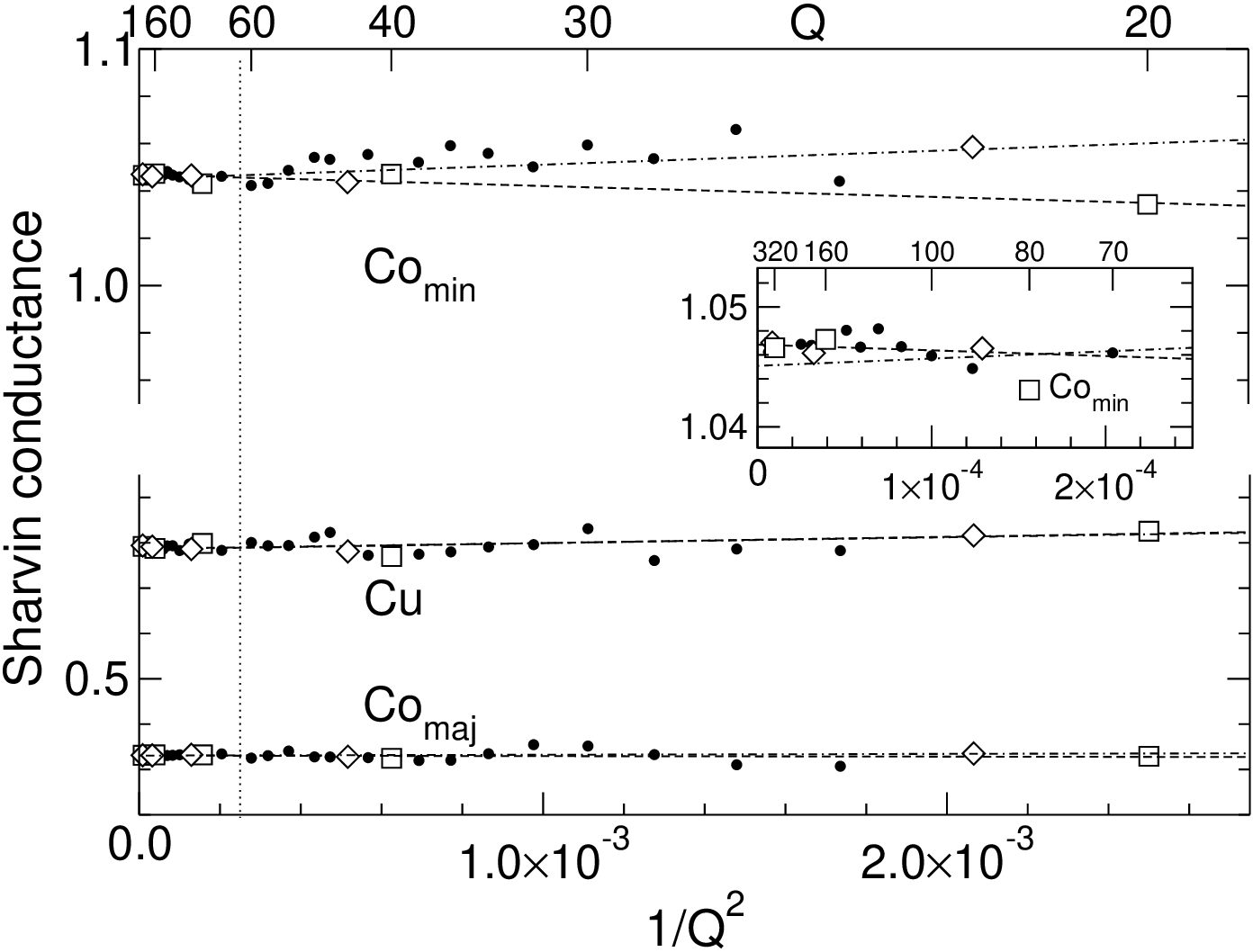}
\caption{Sharvin conductance $G^{\sigma }(111)$ (in units of
$10^{15} \: \Omega^{-1} \mathrm{m}^{-2}$) for bulk {\em fcc} Cu
and Co (majority and minority spin) plotted as a function of the
normalized area element used in the Brillouin zone summation,
$\Delta^2 {\bf k_{\parallel }} / A_{BZ} = 1/Q^2$. $Q$, the number
of intervals along the reciprocal lattice vector is indicated at
the top of the figure. Squares represent the series ($Q = 20, 40,
80, 160, 320$) least-squares fitted by the dashed line; diamonds,
the series ($Q = 22, 44, 88, 176, 352$) least-squares fitted by
the dash-dotted line. The part of the curve for the Co minority
spin case to the left of the vertical dotted line is shown on an
expanded scale in the inset. An \textit{fcc} lattice constant of
$a=3.614\mathring{A}$ and \textit{spd} basis were used together
with von Barth-Hedin's exchange-correlation potential.}
\label{fig:lead_conv}
\end{figure}

The $\mathbf{k}_{\parallel }$ summation is carried out by sampling,
on a regular mesh, the 2D Brillouin zone (BZ) defined by the (lateral)
translational periodicity perpendicular to $\hat{n}$.
The results of carrying out this BZ summation are shown in Fig.~\ref{fig:lead_conv}
where $G^{\sigma }(\hat{n})$ is plotted as a function of
$\Delta^{2}k_{\parallel }/A_{BZ}$, the normalized area element per
$\mathbf{k}_{\parallel }$-point for bulk {\em fcc} Cu and for the
majority and minority spins of bulk \textit{fcc} Co.
When the 2D-BZ reciprocal lattice vectors are each divided into $Q$
intervals, then $\Delta^{2}k_{\parallel } / A_{BZ} = 1/Q^2$.
It can be seen that the Sharvin conductance is converged to about 1\% if
$3600=60\times 60$ points are used in the complete 2D-BZ and to about 0.2\%
for $102400=320\times 320$ sampling points.
The worst case is for the minority spin of Co which has a complex
multi-sheeted Fermi surface.
To see if there are any simple underlying trends in the convergence, we
repeatedly bisect the intervals used in the BZ summation starting with
$Q=20$ and $Q=22$, shown in the figure as squares and diamonds, respectively
and least-squares fitted with the dashed and dash-dotted lines.
The convergence is fairly uniform but not very systematic indicating that
the summation is limited by fine structure in the integrand at the smallest
length scale studied which can only be resolved by increasingly fine sampling.
Thus there is nothing to be gained by developing more sophisticated
interpolation schemes and when we introduce disorder in
Section~\ref{ssec:IntDis}, this will be even more so.
However, in the following we will see that the level of convergence we can
achieve with discrete sampling is quite adequate and not a limiting step
in the whole procedure.

The calculations shown in the figure were performed using an
$spd$-basis, for an \textit{fcc} lattice constant $a=3.614 \,
\mathring{A}$ corresponding to the experimental volume of bulk
(\textit{fcc}) Cu and using the exchange-correlation potential
calculated and parameterized by von Barth and
Hedin.\cite{vonBarth:jpc72} For convenience, and to avoid repetition,
we will refer to this in the following as a ``standard''
configuration.  The converged values are given (underlined) in
Table~\ref{tab:A} together with values calculated using an
\textit{fcc} lattice constant $a=3.549 \, \mathring{A}$ corresponding
to the volume of bulk \textit{hcp} Co.\cite{fn:WS_radii} Because we
shall be studying Cu/Co interfaces where the volume per atom is not
known very precisely from experiment, we will want to estimate the
variation that can be expected when different but equally reasonable
lattice constants are used. The increase of 3.4\% (from 0.558 to
$0.577 \times 10^{15} \: \Omega^{-1} \mathrm{m}^{-2}$) observed for Cu
can be attributed to the increased areal density of Cu atoms,
$(3.614/3.549)^2$ corresponding to $\sim 3.7\%$. The Table also
contains the corresponding results obtained with an
\textit{spdf}-basis.  To the numerical accuracy shown, there is no
difference between the \textit{spd} and \textit{spdf} case for Cu.

For Co majority spin states, there is a 4\% decrease in the
conductance on going from an \textit{spd} to an \textit{spdf} basis.
For a lattice constant $a=3.614 \, \mathring{A}$, the magnetization is
1.684$\, \mu_B$/atom for an \textit{spd}- and
1.648$\, \mu_B$/atom for an \emph{spdf} basis corresponding, respectively,
to $n_{\rm maj}=$ 5.342 and 5.324 electrons in the majority spin bands.
Since all five (nominal) majority-spin $d$ bands are full there are
0.342 and 0.324 electrons in the free-electron-like $sp$ band. In a free
electron
picture the ratio of the projection of the spherical Fermi surfaces is
$(0.324/0.342)^{2/3}=0.96$, thus explaining the observed numerical
result.

\begin{table}[btp]
\begin{ruledtabular}
\begin{tabular}{lcllll}
         &           &        &  & \multicolumn{2}{c} {$G^{\sigma }(111)$}\\ \cline{5-6}
         & $a$(\AA ) & basis  & $n_{\sigma}$ & present calc. &
                                                    Schep\footnotemark[1] \\ \hline
Copper   & 3.549     & $spd$  & 5.5   & 0.577(0.577,0.577)       & 0.57 \\
         & 3.549     & $spdf$ & 5.5   & 0.577(0.577)             & ---- \\
         & 3.614     & $spd$  & 5.5   & \underline{0.558}(0.559) & 0.55 \\
         & 3.614     & $spdf$ & 5.5   & 0.558(0.558)             & 0.55 \\
\\
Cobalt   & 3.549     & $spd$  & 5.323 & 0.469(0.459,0.467)       & 0.45 \\
majority & 3.549     & $spdf$ & 5.304 & 0.449(0.440)             & 0.43 \\
         & 3.614     & $spd$  & 5.342 & \underline{0.466}(0.457) & 0.45 \\
         & 3.614     & $spdf$ & 5.324 & 0.448(0.439)             & ---- \\
\\
Cobalt   & 3.549     & $spd$  & 3.677 & 1.082(1.081,1.082)       & 1.10 \\
minority & 3.549     & $spdf$ & 3.696 & 1.120(1.125)             & 1.13 \\
         & 3.614     & $spd$  & 3.658 & \underline{1.046}(1.047) & 1.06 \\
         & 3.614     & $spdf$ & 3.676 & 1.074(1.079)             & ---- \\

\end{tabular}
\end{ruledtabular}
\footnotetext[1]{Ref.\onlinecite{Schep:prb98} }
\caption[Tab1]{
The Sharvin conductances per spin (in units of $10^{15} \:
\Omega^{-1} \mathrm{m}^{-2}$) in the (111) direction for
\emph{fcc} Cu and Co using the experimental volumes of Cu and Co.
The underlined numbers are the converged values discussed in relation
to Fig.~\ref{fig:lead_conv}.
Most of the results were obtained with von Barth-Hedin's
exchange-correlation potential while the results in brackets are
for Perdew-Zunger (PZ) and Vosko-Wilk-Nusair (VWN) parameterizations,
respectively. Where a single number is given in brackets, it means
that PZ and VWN potentials yield identical results to the accuracy
given.
The corresponding results of Schep {\em et al.} are
given in the last column. The number of electrons with spin
$\sigma$ is given in the fourth column. }
\label{tab:A}
\end{table}

The Co majority-spin conductance scarcely changes with changing
lattice constant, however. The origin of this behaviour lies in the
volume dependence of the magnetic moment. When the lattice constant is
decreased, the $d$ bands broaden and the magnetic moment decreases from
1.684 to 1.646$\, \mu_B$/atom in the $spd$ case with a corresponding
decrease of the occupancy of the $sp$ band from 0.342 to 0.323
majority-spin electrons. The corresponding 4\% decrease in conductance
is almost perfectly compensated by the increased areal density of atoms
so there is no net change.
For the minority-spin conductance, the same factors play a role but
now the $d$ bands are only partly filled. This results in complex
Fermi surfaces for which simple estimates cannot be made. In this
case recourse must be made to full band structure calculations.
We return to this in Sect.~\ref{ssec:OrdInt}.

The calculations presented so far were carried out using the
exchange-correlation potential calculated and parameterized by
von Barth and Hedin.\cite{vonBarth:jpc72}
This is only one of a number of potentials we could have used, none of
which is clearly better than the others in describing the ground state
properties of magnetic materials.
To gauge the uncertainty arising from this arbitrary choice, a number of
calculations were carried out using the potentials given by
Perdew-Zunger\cite{Perdew:prb81} and Vosko-Wilk-Nusair\cite{Vosko:cjp80}
and the results are given in brackets in the Table. Using different
exchange-correlation potentials leads to variation in the conductances
of the order of 1 or 2 percent.

A different (but equivalent) approach was adopted by
Schep \emph{et al.}\cite{Schep:prl95,Schep:prb98} to
the determination of the Sharvin conductances for the same systems using
conventional first-principles LMTO-ASA bulk electronic band structures,
i.e. using $\varepsilon _{i}(\mathbf{k})$ rather than
$k_{\mu}(\varepsilon =\varepsilon _{F},\mathbf{k}_{\parallel })$ as used
here. He expressed
the Sharvin conductance as a projection of the Fermi surface onto a plane
perpendicular to the transport direction and calculated the areas using
a suitably modified 3D-BZ integration scheme. His results are also given in
Table~\ref{tab:A} and are as consistent with our present values as can be
expected when using two entirely different computer codes.

In determining the conductance of the leads, the BZ summation does not
present a problem. The uncertainties arising from small variations in
the atomic volumes, from incompleteness of the basis and from the choice
of LDA parameterization are of comparable size. The MTO-AS approximation
can be systematically improved but only at substantial computational
cost. Since there is currently no way to systematically improve upon the
LDA we identify it and the lack of knowledge of the atomic structure
as limiting factors in studying transport from first principles. Though
the atomic structures could be determined theoretically by total energy
minimization, the LDA again presents a barrier since it systematically
underestimates lattice constants of transition metals in particular of
the 3$d$ series. Gradient corrections sometimes yield improvements but
unfortunately not systematically so. We conclude that our knowledge of
and ability to calculate from first principles Fermi surfaces for bulk
magnetic materials such as Fe or Co does not at present justify using
a more accurate but substantially more expensive computational scheme
than the present one.

\subsection{Ordered Interfaces}
\label{ssec:OrdInt}

Cu and Co have slightly different atomic volumes. The equilibrium lattice
constant of Cu is $3.614 \, \mathring{A}$ and of Co $3.549 \, \mathring{A}$,
assuming an {\em fcc} structure. Even in the absence of interface disorder,
the lattice spacing will not be homogeneous and will depend on the lattice
constant of the substrate on which the sample was grown, on the global and
local concentrations of Cu and Co, and on other details of how the structure
was prepared. In principle we could calculate all of this by energy
minimization. However, we judge that the additional effort needed is not
justified by current experimental knowledge. Instead, we content
ourselves with estimating the uncertainty which results from plausible
variations in the (interface) structure by considering two limiting cases
and one intermediate case. In each case an {\em fcc} structure is assumed,
with lattice constants corresponding to
(i) the atomic volume of Cu,
(ii) the atomic volume of Co,
(iii) an intermediate case with arithmetic mean of Cu and Co atomic volumes.
\begin{table}[b]
\begin{ruledtabular}
\begin{tabular}{l D{.}{.}{4} D{.}{.}{5} D{.}{.}{5} D{.}{.}{5}}
\multicolumn{1}{c}{$a (\mathring{A}$)} &
       \multicolumn{2}{c} {3.549} &
                         \multicolumn{1}{c} {3.582} &
                                     \multicolumn{1}{c} {3.614} \\
\hline
\multicolumn{1}{c}{Basis} &
    \multicolumn{1}{c}{\em spdf} &
                      \multicolumn{1}{c}{\em spd} &
                                    \multicolumn{1}{c}{\em spd} &
                                                 \multicolumn{1}{c}{\em spd} \\
\hline
$m_{\rm Cu}$(bulk)  &  0.000      &  0.000        &  0.000      &  0.000     \\
$m_{\rm Cu}$(int-4) &  0.001(1)   &  0.001        &  0.001      &  0.001     \\
$m_{\rm Cu}$(int-3) & -0.001(0)   &  0.000        &  0.000      &  0.000     \\
$m_{\rm Cu}$(int-2) & -0.005(5)   & -0.005(4,5)   & -0.005(4)   & -0.005     \\
$m_{\rm Cu}$(int-1) &  0.002(4)   &  0.004(6,4)   &  0.003(4)   &  0.001(2)  \\
\hline
$m_{\rm Co}$(int+1) & 1.526(490)  &  1.578(45,73) &  1.605(573) &  1.636(01) \\
$m_{\rm Co}$(int+2) & 1.621(597)  &  1.656(35,53) &  1.673(53)  &  1.690(70) \\
$m_{\rm Co}$(int+3) & 1.602(576)  &  1.645(21,41) &  1.662(39)  &  1.680(59) \\
$m_{\rm Co}$(int+4) & 1.610(587)  &  1.649(27,45) &  1.665(45)  &  1.683(62) \\
$m_{\rm Co}$(bulk)  & 1.609(590)  &  1.646(22,42) &  1.667(45)  &  1.684(62) \\
\hline
$G^{\rm maj}(111)$  &  0.409(399) &  0.431(21,29) &  0.433(22)  &  0.434 (24) \\
$G^{\rm min}(111)$  &  0.378(379) &  0.378(80,79) &  0.371(73)  &  0.364 (67)
\end{tabular}
\end{ruledtabular}
\caption{Variation of the layer-resolved magnetic moments (in Bohr
magnetons) for Cu/Co(111) interfaces with basis set and lattice constant.
These results were obtained with von Barth-Hedin's exchange-correlation
potential while the results in brackets, where given, are for
Perdew-Zunger and Vosko-Wilk-Nusair parameterizations, respectively.
In the last two rows, the interface conductances are given in units of
$10^{15} \: \Omega^{-1} \mathrm{m}^{-2}$.}
\label{tab:B}
\end{table}

Our starting point is a self-consistent TB-LMTO SGF calculation\cite{Turek:97}
for the interface embedded between semi-infinite Cu and Co leads whose
potentials and spin-densities were determined self-consistently in separate
``bulk'' calculations. The charge and spin-densities are allowed to vary in
$n_{\rm Cu}$ layers of Cu and $n_{\rm Co}$ layers of Co bounding the interface.
The results of these calculations for Cu/Co(111) interfaces and the three
different lattice constants detailed above are given in Table~\ref{tab:B}
for $n_{\rm Cu}$=4, $n_{\rm Co}$=4. In the Cu layers, only tiny moments are
induced. Only four layers away from the interface on the Co side, the
magnetic moments are seen to be very close to the bulk values. At the
interface, where the $d$-bandwidth is reduced as a result of the lower
coordination number, the moments are {\em suppressed} rather than enhanced.
This occurs because the majority-spin $d$ bands are full and their number
cannot increase. The width of the free-electron like $sp$ band is less
sensitive to the change in coordination and its exchange splitting also
changes less. As a result, there is little change in the $sp$ moment.
When the $d$-bandwidth is reduced, there is conversion of minority- and
majority-spin $sp$ electrons, without loss of the $sp$ moment, to the
minority-spin $d$ band with loss of $d$ moment. This picture is supported
by the full calculations.

Earlier we saw that an $\sim 2\%$ change in lattice constant changed
the bulk magnetic moment of \textit{fcc} Co by $2.3 \%$. The effect of
changing the basis, from \textit{spd} to \textit{spdf}, was similar.
From Table~\ref{tab:B}, the interface moments are seen to behave in a
comparable fashion.
The magnetic moment of the interface Co atoms decreases by $3.7 \%$,
from 1.636$\, \mu_B$/atom for $a=3.614 \, \mathring{A}$
to 1.578$\, \mu_B$/atom for $a=3.549 \, \mathring{A}$ for an $spd$ basis
and decreases from 1.578$\, \mu_B$/atom to 1.526$\, \mu_B$/atom for
an $spdf$ basis for $a=3.549 \, \mathring{A}$, a change of $3.4 \%$.
Thus the $sp$ to $d_{\rm min}$ conversion is enhanced at the interface
by the reduced $d$-bandwidth.

\begin{figure}[tbp]
\includegraphics[scale=0.30,clip=true]{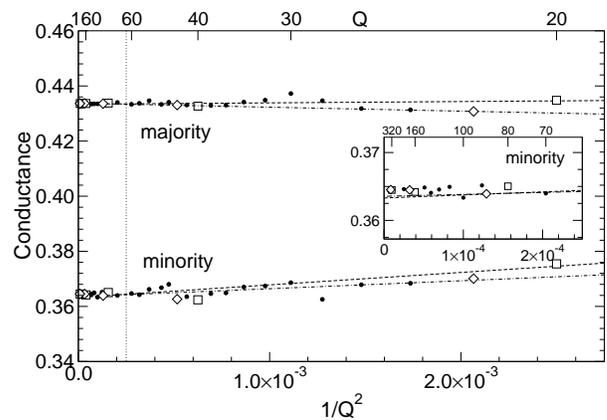}
\caption{Interface conductance $G^{\sigma }(111)$ (in units of
$10^{15} \: \Omega^{-1} \mathrm{m}^{-2}$) for an {\em fcc}
Cu/Co(111) interface for majority and minority spins plotted as a
function of the normalized area element used in the Brillouin zone
summation, $\Delta^2 {\bf k_{\parallel }} / A_{BZ} = 1/Q^2$. $Q$,
the number of intervals along the reciprocal lattice vector is
indicated at the top of the figure. Squares represent the series
($Q = 20, 40, 80, 160, 320$) least-squares fitted by the dashed
line; diamonds, the series ($Q = 22, 44, 88, 176, 352$)
least-squares fitted by the dash-dotted line. The part of the
curve for the Co minority spin case to the left of the vertical
dotted line is shown on an expanded scale in the inset.
A ``standard'' configuration was used.\cite{fn:standard}}
\label{fig:cl_int_conv}
\end{figure}

Once the interface potential has been obtained, the transmission matrix
can be calculated and the BZ summation carried out. The convergence of
this summation, shown in Fig.~\ref{fig:cl_int_conv} for a lattice
constant of $a=3.614 \, \mathring{A}$ and an \textit{spd} basis closely
parallels that seen in Fig.~\ref{fig:lead_conv} and therefore the
k-summation does not represent a limitation in practice. Converged
conductances
\begin{equation}
G^{\sigma }(\hat{n})
= \frac{e^2}{h} \sum_{\mu,\nu,{\bf k_{\parallel}}}
                T_{\mu \nu}^{\sigma } ({\bf k_{\parallel }})
= \frac{e^2}{h} \sum_{\mu,\nu,k_{\parallel}}
               |t_{\mu \nu }^{\sigma }({\bf k_{\parallel }})|^2
\label{eq:TransMat}
\end{equation}
are given in the last two rows of Table~\ref{tab:B}.
Though we will not concern ourselves in this publication with the
application of the formalism we have been developing to a detailed
interpretation of experimental observations, it should be noted that
even a modest spin-dependence of ``bare'' interface conductances
($\sim 20 \%$) can lead to spin-dependent interface resistances
differing by a factor of $\sim 3-5$ once account is taken of the
finiteness of the conductance of the perfect leads using a formula
derived by Schep {\em et al.}\cite{Schep:prb97}
\begin{equation}
R_{A/B}^{\sigma }=\frac{h}{e^{2}}\left[
\frac{1}{\sum T_{\mu \nu }^{\sigma }} -
\frac{1}{2}\left(
\frac{1}{N_{A}^{\sigma }}+\frac{1}{N_{B}^{\sigma }}\right) \right]
\label{eq:R_Schep}
\end{equation}
where $N_A^{\sigma }$ and $N_B^{\sigma }$ are the Sharvin conductances
(see Eq. \eqref{eq:Sharvin}) of the materials A and B forming the
interface, in units of ${e^2 / h}$.

The majority-spin case can be readily understood in terms of the
geometry of the Fermi surfaces of Cu and Co so we begin by discussing
this simple case before examining the more complex minority-spin
channel.

\subsubsection*{Clean Cu/Cu (111) Interface: Majority Spins}

\begin{figure}[tbp]
\includegraphics[scale=0.43,clip=true]{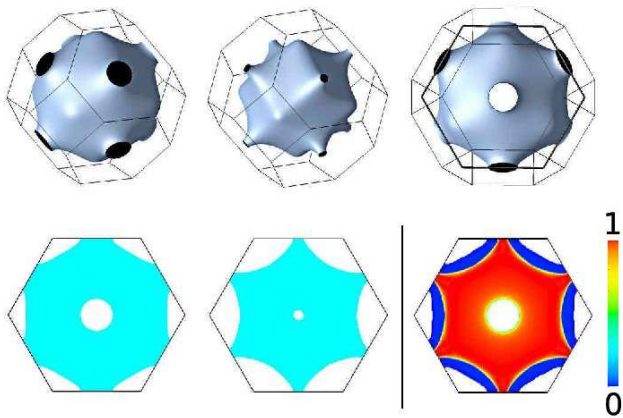}
\caption{ Top row, left-hand panel: Fermi surface (FS) of Cu;
middle panel: majority-spin FS of Co; right-hand panel: Cu FS
viewed along the (111) direction with a projection of the bulk fcc
Brillouin zone (BZ) onto a plane perpendicular to this direction
and of the two dimensional BZ.
Bottom row, left-hand and middle panels: projections onto a plane
perpendicular to the (111) direction of the Cu and majority-spin
Co Fermi surfaces;
right-hand panel: transmission probability for majority-spin states
as a function of transverse crystal momentum, $T({\bf k_{\parallel}})$
for an {\em fcc} Cu/Co(111) interface.
A ``standard'' configuration was used.\cite{fn:standard}
}
\label{fig:CuCo111_maj}
\end{figure}

In the absence of disorder, crystal momentum parallel to the interface
is conserved. If, for a given value of ${\bf k}_{\parallel}$, there
is a propagating state in Cu incident on the interface but none in Co,
then an electron in such a state is completely reflected at the interface.
Conversely, ${\bf k}_{\parallel}$'s for which there is a propagating state
in Co but none in Cu also cannot contribute to the conductance.
To determine the existence of such states, it is sufficient to inspect
projections of the Fermi surfaces of {\em fcc} Cu and majority-spin
Co onto a plane perpendicular to the transport direction ${\hat n}$,
shown in Fig.~\ref{fig:CuCo111_maj} for ${\hat n}=(111)$. The first
feature to note in the figure (left-hand and middle panels) is that per
${\bf k}_{\parallel}$ there is only a single channel with positive group
velocity so that the transmission matrix in \eqref{eq:TransMat} is a
complex number whose modulus squared is a transmission probability with
values between 0 and 1. It is plotted in the
right-hand panel and can be interpreted simply. Regions which are
depicted blue correspond to ${\bf k}_{\parallel}$'s for which there are
propagating states in Cu but none in Co. These states have transmission
probability 0 and are totally reflected. For values of
${\bf k}_{\parallel}$ for which there are propagating states in both
Cu and Co, the transmission probability is very close to one, depicted
red. These states are essentially free electron-like states which have
the same symmetry in both materials and see the interface effectively
as a very low potential step.
Close to the centre of the figure there is an annular region where
there are propagating states in Co but none in Cu so they do not
contribute to the conductance.
Performing the sum in \eqref{eq:TransMat}, we arrive at an interface
conductance of $0.434 \times 10^{15} \: \Omega^{-1} \mathrm{m}^{-2}$
to be compared to the Sharvin conductances given in Table~\ref{tab:A}
for Cu and Co; for $a=3.614 \, \mathring{A}$ and an $spd$ basis these
are, respectively, 0.558 and 0.466 in the same units. The interface
conductance of 0.434 is seen to be essentially the Sharvin conductance
of the majority states of Co reduced because the states closest to the
$\Lambda$-axis (corresponding to the symmetry axis of the figures, the
$\Gamma L$ line in reciprocal space) do not contribute. The explanation
of the 5\% decrease found on going from an $spd$ to an $spdf$ basis,
(0.431 to 0.409), parallels that given for the corresponding change in
the Sharvin conductance of bulk Co (0.469 to 0.449 in Table~\ref{tab:A}).

\subsubsection*{Clean Cu/Cu (111) Interface: Minority Spins}

\begin{figure*}[btp]
\includegraphics[scale=0.95,clip=true]{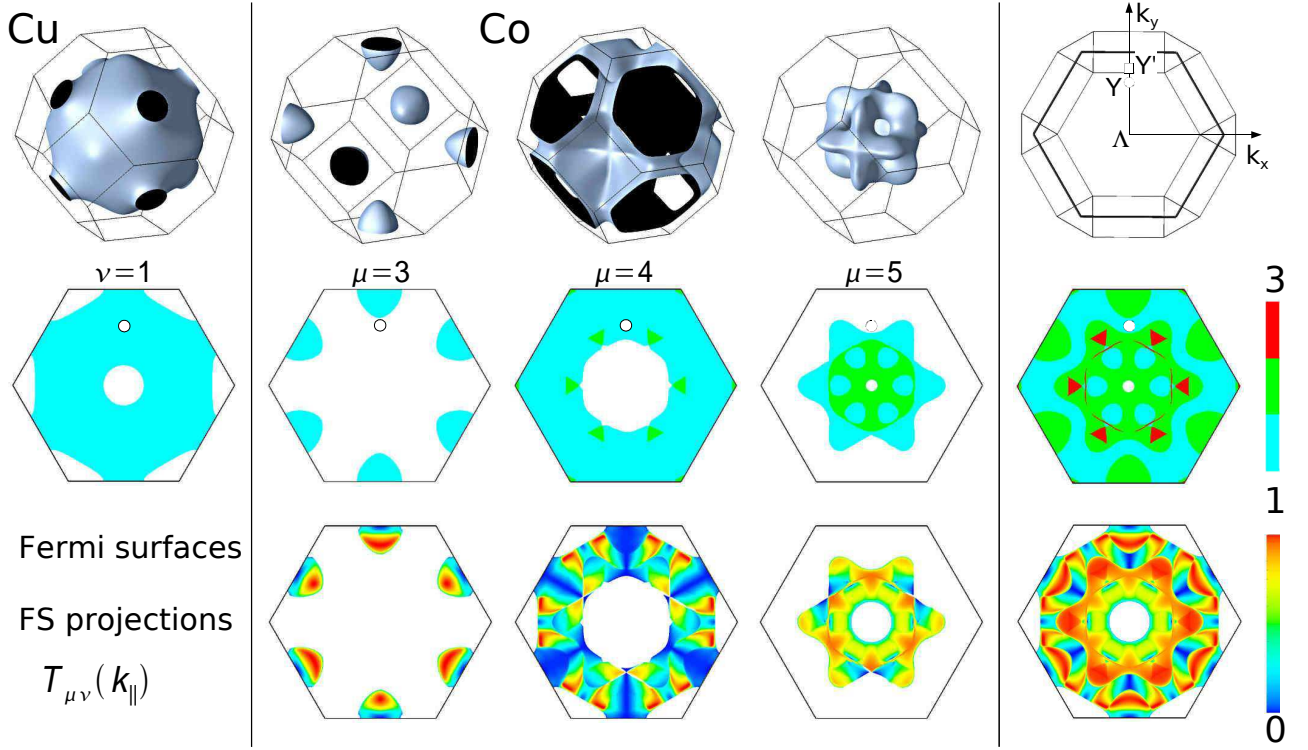}
\caption{ Top row, lefthand panel: Fermi surface (FS) of
\emph{fcc} Cu; middle panels: third, fourth and fifth FS sheets of
minority-spin fcc Co; righthand panel: projection of the bulk fcc
Brillouin zone (BZ) onto a plane perpendicular to the (111)
direction and of the two dimensional BZ.
Middle row: corresponding projections of individual FS sheets and (rhs)
of Co total. The number of propagating states with positive velocity is
colour-coded following the colour bar on the right.
Bottom row: probability $T_{\mu \nu}(\mathbf{k_{\parallel }})$
for a minority-spin state on the single FS sheet of Cu ($\nu = 1$)
to be transmitted through a Cu$|$Co(111) interface into FS sheet
$\mu$ of \emph{fcc} Co as a function of the transverse crystal
momentum $\mathbf{k_{\parallel }}$. The point Y is such that there
are only propagating states in Cu and in the fourth FS sheet of
Co. For the point Y' slightly further away from $\Lambda $ and
indicated by a small open square there is, in addition, a
propagating state in the third FS sheet of Co.
Results are for a ``standard'' configuration.\cite{fn:standard}
 }
\label{fig:CuCo111_min}
\end{figure*}

The minority-spin case is considerably more complex because the
Co minority-spin $d$ bands are only partly filled, resulting in
multiple sheets of Fermi surface. These sheets are shown in
Fig.~\ref{fig:CuCo111_min} together with their projections onto a
plane perpendicular to the (111) transport direction. Compared to
Fig.~\ref{fig:CuCo111_maj}, one difference we immediately notice
is that even single Fermi surface (FS) sheets are not single valued:
for a given ${\bf k}_{\parallel}$ there can be more than one mode
with positive group velocity. The areas depicted green in the
projections of the FS sheets from the fourth and fifth bands are
examples where this occurs.

An electron incident on the interface from the Cu side, with transverse
crystal momentum
${\bf k}_{\parallel}$, is transmitted into a linear combination of
all propagating states with the same ${\bf k}_{\parallel}$ in Co; the
transmission matrix $t_{\mu \nu }^{\sigma }({\bf k_{\parallel }})$ is
in general not square but rectangular. The transmission probabilities
$T_{\mu \nu}({\bf k_{\parallel }})$ are shown in the bottom row of
Fig.~\ref{fig:CuCo111_min}. Because there is only a single incident
state for all ${\bf k}_{\parallel}$, the maximum transmission
probability is one. Comparison of the total minority-spin
transmission probability $T_{\mathcal{LR}}({\bf k_{\parallel}})$
(Fig.~\ref{fig:CuCo111_min}, bottom right-hand panel) with the
corresponding majority-spin quantity (right-hand panel of
Fig.~\ref{fig:CuCo111_maj}) strikingly illustrates the spin-dependence
of the interface scattering, much more so than the integrated
quantities might have led us to expect; the interface conductances,
0.364 and $0.434 \times 10^{15} \: \Omega^{-1} \mathrm{m}^{-2}$
from Table~\ref{tab:B}, differ by only $\sim 20\%$.

Three factors contribute to the large ${\bf k}_{\parallel}$-dependence
of the transmission probability: first and foremost, the complexity of
the Fermi surface of both materials but especially of the minority spin
of Co; secondly and inextricably linked with the first because of the
relationship $\hbar\upsilon_{\bf k}={\nabla}_k \varepsilon({\bf{k}})$,
the mismatch of the Fermi velocities of the states on either side of
the interface. Thirdly, the orbital character of the states $\mu$ and
$\nu$ which varies strongly over the Fermi surface and gives rise to
large matrix element effects.

The great complexity of transition metal Fermi surfaces, clear from the
figure and well-documented in standard textbooks, is not amenable to
simple analytical treatment and has more often than not been neglected
in theoretical transport studies. Nevertheless, as illustrated
particularly well by the ballistic limit,\cite{Schep:prl95,Schep:prb98}
spin-dependent band structure effects have been shown to lead to
magnetoresistance ratios comparable to what are observed experimentally
in the current-perpendicular-to-plane (CPP) measuring configuration and
cannot be simply ignored in any quantitative discussion.
Most attempts to take into account contributions of the $d$ states to
electronic transport do so by mapping the five $d$ bands onto a single
tight-binding or free-electron band with a large effective mass.

Fermi surface topology alone cannot explain all aspects of the
tranmission coefficients seen in Fig.~\ref{fig:CuCo111_min}. For
example, there are values of ${\bf k_{\parallel}}$, such as that
labelled $Y$ in the figure, for which propagating solutions exist on
both sides of the interface yet the transmission probability is zero.
This can be understood as follows. 
At ${\bf k_{\parallel}} =Y$, the propagating states in Cu have \{$s,
p_y, p_z, d_{yz}, d_{3z^2 - r^2}, d_{x^2 - y^2}$\} character (assuming
the choice of in-plane axes as illustrated in the top righthand panel
of Fig.~\ref{fig:CuCo111_min}) and are even with respect to
reflection in the plane formed by the $y$-axis and the transport
direction perpendicular to the (111) plane which we choose to be the
$z$-axis.  For this ${\bf k_{\parallel}}$ the only propagating state
in Co is in the fourth band. It has \{$p_x, d_{xy}, d_{xz}$\}
character which is odd with respect to reflection in the $yz$ plane.
Consequently, the corresponding hopping matrix elements in the
Hamiltonian (and in the Green's function) vanish and the transmission
is zero.

Along the $k_y$ axis the symmetry of the states in Cu and those in the
fourth band of Co remain the same and the transmission is seen to vanish
for all values of $k_y$.
However, at points further away from $\Lambda$, we encounter states in
the third band of Co which have even character whose matrix elements do
not vanish by symmetry and we see substantial transmission probabilities.
Similarly, for points closer to $\Lambda$, there are states in the
fifth band of Co with even character whose matrix elements also do not
vanish and again the transmission probability is substantial. Because
it is obtained by superposition of transmission probabilities from Cu
into the third, fourth and fifth sheets of the Co FS, the end result,
though it may appear very complicated, can be straightforwardly
analysed in this manner k-point by k-point.

Though the underlying lattice symmetry is only threefold, the Fermi
surface projections shown in Fig.~\ref{fig:CuCo111_min} have six-fold
rotational symmetry about the line $\Lambda$ because the bulk $fcc$
structure has inversion symmetry (and time-reversal symmetry). The
interface breaks the inversion symmetry so $T_{\mu \nu}({\bf
  k_{\parallel }})$ has only threefold rotation symmetry for the
individual FS sheets. However, in-plane inversion symmetry is
recovered for the total transmission probability $T_{\mathcal{LR}}( -
{\bf k_{\parallel}}) =T_{\mathcal{LR}}( {\bf k_{\parallel}})$ which
has full sixfold symmetry. This follows from the time-reversal
symmetry and is proven in Appendix~\ref{sec:symmetry}.

\subsection{Interface Disorder}
\label{ssec:IntDis}

Instructive though the study of perfect interfaces may be in gaining
an understanding of the role electronic structure mismatch may play
in determining giant magnetoresistive effects, all measurements are
made on devices which contain disorder, mostly in the diffusive regime.
Because there is little information available from experiment about
the nature of this disorder, it is very important to be able to model
it in a flexible manner, introducing a minimum of free parameters. To
model interfaces between materials with different lattice constants
and disorder, we use the lateral supercells described in section
\ref{ssec:Supercells}. Since this approach is formally only valid if
sufficiently large supercells are used, we begin by studying how the
interface conductance depends on the lateral supercell size.

\begin{figure}[tbp]
\includegraphics[scale=0.35,clip=true]{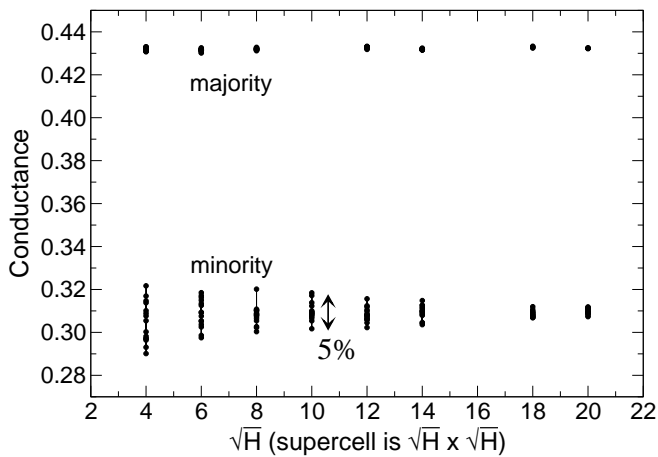}
\caption{Interface conductance (in units of $10^{15} \:
\Omega^{-1} \mathrm{m}^{-2}$) for a disordered Cu/Co (111)
interface modelled as 2ML of 50\%-50\% alloy in a $\sqrt{H} \times
\sqrt{H}$ lateral supercell as a function of 
$\sqrt{H}$. The results are given for different randomly generated
configurations of disorder (15 for minority spin, 5 for majority
spin).
Results are for a ``standard'' configuration.\cite{fn:standard} }
\label{fig:config_av}
\end{figure}

To perform fully self-consistent calculations for a number of large
lateral supercells and for different configurations of disorder would
be prohibitively expensive. Fortunately, the coherent potential
approximation (CPA) is a very efficient way of calculating charge and
spin densities for a substitutional disordered $A_x B_{1-x}$ alloy
with an expense comparable to that required for an ordered system with
a minimal unit cell.\cite{Soven:pr67} The output from such
a calculation are atomic sphere potentials for the two sites,
$\upsilon_A$ and $\upsilon_B$. The layer CPA approximation generalizes
this to allow the concentration to vary from one layer to the next.
\cite{Turek:97}

Once $\upsilon_A$ and $\upsilon_B$ have been calculated for some
concentration $x$, an $H = H_1 \times H_2$ lateral supercell is
constructed in which the potentials are distributed at random,
maintaining the concentration for which they were self-consistently
calculated. The conductances calculated for $4 \leq \sqrt{H} \leq 20 $ are shown in
Fig.~\ref{fig:config_av} for a Cu/Cu(111) interface in which the Cu and
the Co layers forming the interface are totally mixed to give two layers
of 50\%-50\% interface alloy. For each value of $H$, the results for a
number of different randomly generated disorder configurations are shown
(20 for minority, 5 for minority spin). The sample to sample variation
is largest for the minority spin case, ranging from $\pm 5 \%$ for a
modest $4 \times 4$ unit cell and decreasing to less than $\pm 1 \%$
for a $20 \times 20$ unit cell. For $\sqrt{H} \sim 10$, the spread in
minority spin conductances is $\sim 5 \%$ which is comparable to the
typical uncertainty we associated with the LDA error, the uncertainty
in lattice constants or the error incurred by using the ASA.

Comparing now the conductances without and with disorder, we see that
interface disorder has virtually no effect on the majority spin channel
(0.434 versus $ 0.432 \times 10^{15} \: \Omega^{-1} \mathrm{m}^{-2}$)
which is a consequence of the great similarity of the Cu and Co
majority spin potentials and electronic structures. However, in the
minority-spin channel the effect
(0.364 versus $ 0.31 \times 10^{15} \: \Omega^{-1} \mathrm{m}^{-2}$) is
much larger. As noted in the context of \eqref{eq:R_Schep}, a relatively
small change in the interface transmission can lead to a large change in
the interface resistance when account is taken of the finite conductance
of the leads. We will return to the consequences for the spin-dependent
interface resistance after completing the study of the interface
transmission on which it is based.

When disorder is modelled in lateral supercells, the transmission
probabilities can be classified as being {\em specular} or {\em diffuse}
depending upon whether or not transverse momentum is
conserved.\cite{Bruno:jmmm99,Drchal:prb02}
In the presence of interface disorder, the conductance per unit area
can be expressed as
\begin{align}
G &= G_{s} + G_{d}    \nonumber\\
  &= \frac{e^2}{h} \sum_{\substack{ \mu \nu  \\
                                  {\bf k}_{\parallel}  }}
T_{\mu \nu } ({\bf k}_{\parallel}, {\bf k}_{\parallel})
  + \frac{e^2}{h} \sum_{\substack{ \mu \nu  \\
                                  {\bf k}_{\parallel} \neq {\bf k}_{\parallel}^{\prime }  }}
T_{\mu \nu } ({\bf k}_{\parallel }, {\bf k}_{\parallel }^{\prime })
\label{eq:sd}
\end{align}%
where ${\bf k}_{\parallel }$ and ${\bf k}_{\parallel }^{\prime }$ belong to the
two dimensional Brillouin zone for ($1\times 1$) translational periodicity
and
$T_{\mu \nu }({\bf k_{\parallel }, k_{\parallel }^{\prime }})=
 t_{\mu \nu }({\bf k_{\parallel }, k_{\parallel }^{\prime }})
 t_{\mu \nu }^{\dagger }({\bf k_{\parallel }, k_{\parallel }^{\prime }})$.
The transmission matrix elements between two Bloch states with the
same ${\bf k_{\parallel }}$ are defined to be specular, those between
scattering states with different ${\bf k_{\parallel }}$ as being diffuse.
In the absence of interface disorder, there is by definition only a specular
component.

\subsubsection*{Dependence of interface conductance on alloy concentration}

\begin{figure}[tbp]
\includegraphics[scale=0.7,clip=true]{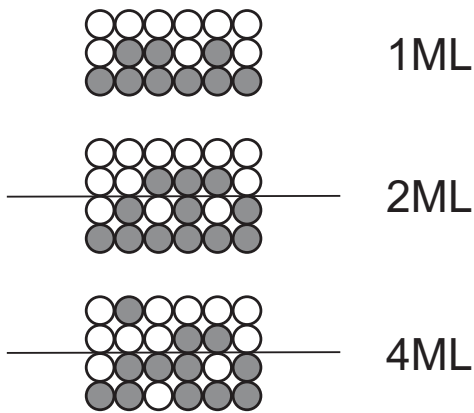}
\caption{Illustration of 3 different models of interface disorder
considered.
Top (1ML): disorder is modelled using one monolayer (ML) of [Cu$_{1-x}$Co${_x}$]
alloy between Cu and Co leads, denoted as Cu$[$Cu$_{1-x}$Co$_{x}]$Co.
Middle (2ML): disorder modelled in two MLs as
Cu$[$Cu$_{1-x}$Co$_{x}|$Cu${_x}$Co$_{1-x}]$Co.
Bottom (4ML): starting from the 2 ML disorder case, 1/3 of the
concentration $x$ of impurity atoms is transferred to the next layer
resulting in disorder in four MLs:
Cu$[$Cu$_{1-\frac{x}{3}}$Co$_\frac{x}{3} | $Cu$_{1-\frac{2x}{3}}$Co$_{\frac{2x}{3}} |
$Cu$_{\frac{2x}{3}}$Co$_{1-\frac{2x}{3}}|$Cu$_\frac{x}{3}$Co$_{1-\frac{x}{3}}]$Co.
}
\label{fig:disord}
\end{figure}

\begin{figure}[t]
\includegraphics[scale=0.35,clip=true]{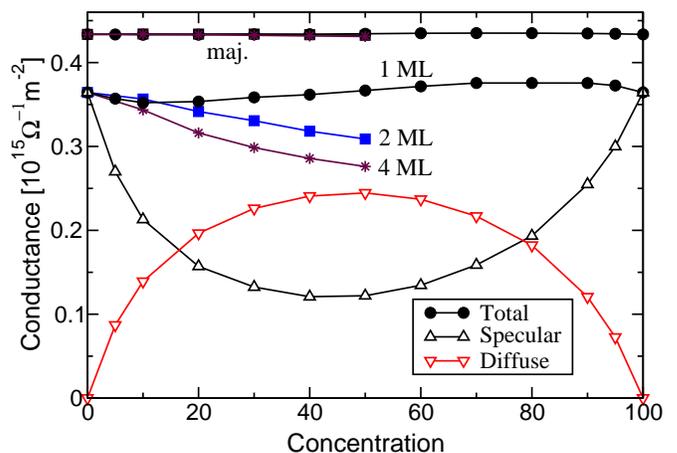}
\caption{Interface conductance of a disordered Cu/Co (111)
interface with disorder modelled in a $20 \times 20$ lateral
supercell as a function of the alloy concentration $x$. Results
are shown for the three different models described in
Fig.~\ref{fig:disord} with disorder in 1, 2 or 4 MLs. Only a
single disorder configuration was used and the size of the symbols
corresponds to the spread in values found for this supercell size
in Fig.~\ref{fig:config_av}. 
For 1ML, the total conductance
is resolved into specular and diffuse components (see the legend). 
Results are for a ``standard'' configuration.\cite{fn:standard}}
\label{fig:spec_diff}
\end{figure}

The results in Fig.~\ref{fig:config_av} were obtained for a structural
model of the Co/Cu(111) interface consisting of two monolayers (2ML)
of 50\%-50\% alloy that was derived from X-ray\cite{Henry:prb96}
NMR\cite{deGronckel:prb91,Meny:prb92} and magnetic
EXAFS\cite{Kapusta:jac99} studies.  Though the most plausible model
there is at present, it contains large uncertainties. This makes it
important to explore the consequences of varying the parameters
defining the model. To do so, we calculate the conductance using 20x20
lateral supercells as a function of alloy concentration for models in
which the disorder is confined to one, two or four monolayers. The
three models are defined in Fig.~\ref{fig:disord}. From the results
shown in Fig.~\ref{fig:spec_diff}, it can be seen that the interface
transmission for majority-spin electrons depends only very weakly on
alloy concentration and its spatial distribution. The results for the
1ML, 2ML and 4ML models cannot be distinguished on the scale of the
figure. When the conductance is decomposed using \eqref{eq:sd}, the
diffuse component is found to be very small. Therefore, only the
results for the minority-spin case need be examined in any detail.

We start by varying the alloy concentration over the full concentration
range (0-100\%) for a disordered monolayer. The magnitude of resulting variation
in transmission is limited ($\sim 7\%$) but exceeds the statistical
spread between different configurations of disorder, which according
to Fig.~\ref{fig:config_av} is less than $\pm 1 \%$.
On adding Co to a layer of Cu, the transmission decreases, reaches a
minimum for 10\% Co, then increases monotonically up to 70-90\% region
where the transmission is \emph{higher} than for a clean
interface.\cite{fn:xia01} 100\% Co represents a clean interface again,
so this limit must yield the same transmission as 0\% Co.

The variation can be examined in terms of the specular and diffuse
components defined in \eqref{eq:sd}. From Fig.~\ref{fig:spec_diff},
it can be seen that, for the minority spin channel, the diffuse
scattering by Co impurity atoms in Cu is stronger than that by Cu
impurity atoms in Co. However, the specular scattering is also more
strongly reduced by Cu in Co than by Co in Cu. The two effects largely
cancel resulting in the observed undulatory total transmission as a
function of the alloy concentration seen in the figure. The diffuse
scattering has a maximum close to a 50\%-50\% alloy concentration where
its contribution to the conductance is almost twice as large as from
the specular scattering. While the conductance as such is scarcely
effected, the strong diffuse scattering will play an important role in
destroying the phase coherence of the electrons. Ultimately, this will
be the physics underlying the so-called ``two current series resistor''
(2CSR) model.\cite{Zhang:jap91,Lee:jmmm93,Valet:prb93}

If the disorder extends over more than a monolayer, then modelling the
interface as several layers of homogeneous alloy is not obviously
realistic. Instead, one might expect the layers closest to the interface
to be most strongly mixed, the amount of mixing decreasing with the
separation from the interface. A simple way to model this is to take
two interface layers, one Cu and one Co, and to mix them in varying
degrees. Denoting this Cu$|$Co interface as
Cu$[$Cu$_{1-x}$Co$_{x}|$Cu${_x}$Co$_{1-x}]$Co
we consider $0 \le x \le 0.5$ {\em i.e.}, the Cu concentration decreases
monotonically from left to right. The calculated interface transmission
is seen to essentially interpolate linearly the results obtained
previously for the clean ($x=0$) and disordered ($x=0.5$) cases.

A slightly more elaborate model can be constructed from the 2ML model by
distributing the $x$ impurity atoms so that $2x/3$ are in the interface
layer while $x/3$ are to be found further from the original interface,
in the following layer. This results in the concentration profile
Cu$[$Cu$_{1-\frac{x}{3}}$Co$_\frac{x}{3} | $Cu$_{1-\frac{2x}{3}}$Co$_{\frac{2x}{3}} |
$Cu$_{\frac{2x}{3}}$Co$_{1-\frac{2x}{3}}|$Cu$_\frac{x}{3}$Co$_{1-\frac{x}{3}}]$Co.
$x=0$ corresponds to a completely ordered interface while the maximum
value $x$ can have so that the concentration decreases from left to
right monotonically is 75\%.
This relatively small redistribution of intermixed atoms is seen to
reduce the transmission by 15\% for $x=0.5$. Even for very good metals,
relatively opaque interfaces can result when the electronic structure
on either side have different characters. In such situations, disorder
can influence the transmission strongly even reducing the interface
resistance very substantially.\cite{Xia:prb01}
A detailed analysis of the different contributions to the interface
scattering in the 2ML and 4ML cases will be given in a separate
publication.

\subsection{Analysis of Interface Disorder Scattering}
\label{ssec:DefSca}

\begin{figure}[b]
\includegraphics[scale=1.0,angle=0,clip=true]{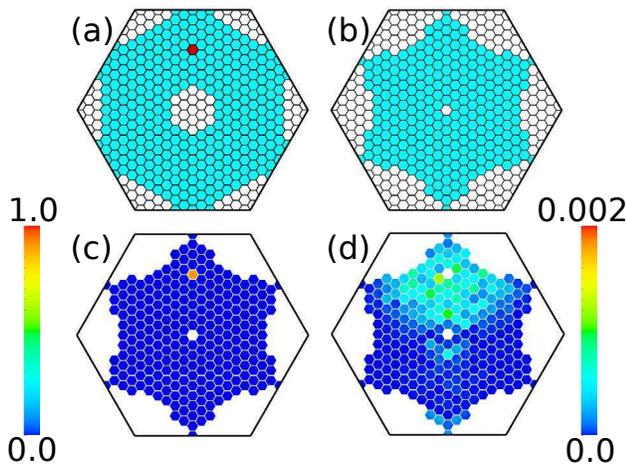}
\caption{Fermi surface projections of majority-spin {\em fcc} Cu
(a) and Co (b) derived from a single k-point using a $20 \times
20$ lateral supercell. The dark red point in the Cu Fermi surface
projection corresponds to the point $Y$ in the top righthand panel
of Fig.~\ref{fig:CuCo111_min}. $T(Y,\bf{k}_{\parallel}^{\prime })$
is shown in (c), and in (d) magnified by a factor 500 where the
ballistic component $T(Y,\bf{k}_{\parallel}^{\prime }=Y)$ is
indicated by a white point because its value goes off the scale.
Results are for a ``standard'' configuration\cite{fn:standard} 
and averaged over 5 different configurations of disorder.}
\label{fig:dis_scatt_maj}
\end{figure}

\begin{figure}[b]
\includegraphics[scale=1.0,angle=0,clip=true]{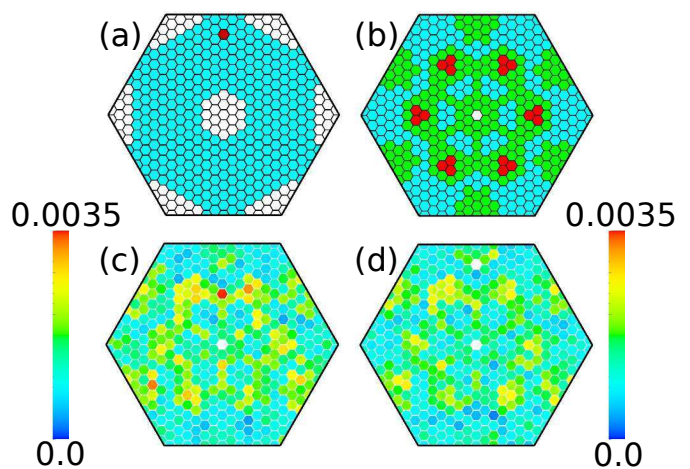}
\caption{Fermi surface projections of minority-spin {\em fcc} Cu
(a) and Co (b) derived from a single k-point using a $20 \times
20$ lateral supercell. The dark red point in the Cu Fermi surface
projection corresponds to the point $Y^{\prime }$ in the top
righthand panel of Fig.~\ref{fig:CuCo111_min}. (c)
$T(Y,\mathbf{{k}_{\parallel }^{\prime })}$ and (d) $T(Y^{\prime
},\mathbf{{k}_{\parallel }^{\prime })}$ calculated using 20
different disorder configurations; the ballistic component
$T(Y^{\prime },\mathbf{{k}_{\parallel }^{\prime }}=Y^{\prime })$
is indicated by a white point because its value goes off scale.
Results are for a ``standard'' configuration,\cite{fn:standard} 
20 different configurations of disorder were used.}
\label{fig:dis_scatt_min}
\end{figure}

The scattering induced by two layers of 50\%-50\% alloy is illustrated
in Fig.~\ref{fig:dis_scatt_maj} and Fig.~\ref{fig:dis_scatt_min} for
the majority and minority spins, respectively, of a Cu/Co(111)
interface.  Calculations were performed for the single
$\bf{k}_{\parallel}^{\mathbb{S}}$ point, $\Gamma $, and a $20 \times
20$ lateral supercell 
equivalent to using a $1 \times 1$ interface
cell and k-space sampling with $20 \times 20$ points in the
corresponding BZ. Disorder averaging was carried out using 5 (for
majority spin) or 20 (for minority spin) disorder configurations generated
randomly.

Figs.~\ref{fig:dis_scatt_maj}(a) and \ref{fig:dis_scatt_maj}(b) show
the majority-spin Fermi surface projections of {\em fcc} Cu and Co,
respectively, obtained from ``unfolding'' the supercell calculation.
The coarse $20 \times 20$ grid is seen to yield a good representation
of the detailed Fermi surface projections shown in
Fig.~\ref{fig:CuCo111_maj}.  $T(\bf{k}_{\parallel
},\bf{k}_{\parallel}^{\prime })$ is shown in
Fig.~\ref{fig:dis_scatt_maj}(c) for ${\bf k_{\parallel}} = Y$ on the
$k_y$ axis in Fig.~\ref{fig:CuCo111_min}. Specular scattering
dominates with $T( {\bf k}_{\parallel} = Y, {\bf
  k}_{\parallel}^{\prime} = Y) = 0.93$.  The diffuse scattering is so
weak that nothing can be seen on a scale of {\em T} from 0 to 1. To
render it visible, a magnification by a factor 500 is needed,
Fig.~\ref{fig:dis_scatt_maj}(d). The total diffuse scattering,
$T_{d}(Y) = \sum_{ k_{\parallel }^{\prime} \neq k_{\parallel } }
T({\bf k_{\parallel }}=Y, {\bf k_{\parallel }^{\prime}} \neq Y) =
0.04$ can be seen from the 
figure to be made up of contributions of $T
\sim 0.0004$ from roughly a quarter 
of the BZ (100 ${\bf
  k_{\parallel}}$ points) centred on ${\bf k_{\parallel }} = Y$. The
total transmission, $T_{total}= T_{s} + T_{d} = 0.93 + 0.04 = 0.97$,
compared to a transmission of 0.99 in the absence of disorder.  
Similar results were obtained for other $\mathbf{k}_{||}$ points.
In the majority case, there is thus a strong specular peak surrounded
by a weak diffuse background.

The minority-spin Fermi surface projections of {\em fcc} Cu and Co are
shown in Figs.~\ref{fig:dis_scatt_min}(a) and \ref{fig:dis_scatt_min}(b),
respectively. Compared to the corresponding panels in
Fig.~\ref{fig:CuCo111_min}, the $20 \times 20$ point representation is
seen to be sufficient to resolve the individual Fermi surface sheets of
Co. To study the effect of interface disorder, we consider scattering
out of two different ${\bf k_{\parallel }}$s in Cu
(Figs.~\ref{fig:dis_scatt_min}(c) and (d)). The first thing to
note is the similarity of both transmission plots  to the projected FS
of Co, Fig.~\ref{fig:dis_scatt_min}(b), suggesting very strong diffusive
scattering proportional to the density of available final states.

The first case we consider is where ${\bf k_{\parallel}}=Y$ for
which the transmission was zero as a result of the symmetry of the
states along the $k_y$ axis in the absence of disorder.
$T({\bf k_{\parallel }}= Y,{\bf k_{\parallel}^{\prime }})$ is shown in
Fig.~\ref{fig:dis_scatt_maj}(c). By contrast with the majority-spin
case just examined, there is now scattering to all other k-points in
the 2D BZ,
$\sum_{     k_{\parallel }^{\prime} \neq k_{\parallel }     }
T({\bf k_{\parallel }}=Y, {\bf k_{\parallel }^{\prime}} \neq Y) = 0.58$
while $T(Y,Y)$ has only increased from 0.00 in the clean case, to 0.01
in the presence of disorder.
The effect of disorder is to increase the total transmission,
$T_{total}(Y)= \sum_{ k_{\parallel }^{\prime}  }
T({\bf k_{\parallel }}=Y, {\bf k_{\parallel }^{\prime}})$ from
0.00 to $T_{s}(Y) + T_{d}(Y) = 0.01 + 0.58 = 0.59$; for states
which were originally strongly reflected, disorder {\em increases} the
transmission.

The second case we consider is that of a k-point slightly further away
from the origin $\Lambda$ along the $k_y$ axis which had a high
transmission, $T(Y^{\prime})= 0.98$, in the absence of disorder. For
this k-point, $T({\bf k_{\parallel }}= Y^{\prime},{\bf
  k_{\parallel}^{\prime }})$, shown in
Fig.~\ref{fig:dis_scatt_maj}(d), looks very similar to
Fig.~\ref{fig:dis_scatt_maj}(c). There is strong diffuse scattering
with $\sum_{ k_{\parallel }^{\prime} \neq k_{\parallel } } T({\bf
  k_{\parallel }}=Y^{\prime}, {\bf k_{\parallel }^{\prime}} \neq
Y^{\prime}) = 0.54$ while $T(Y^{\prime},Y^{\prime})$ has been
drastically decreased from 0.98 in the clean case, to 0.06 as a result
of disorder. The total transmission, $T_{total}(Y^{\prime}) =
T_{s}(Y^{\prime}) + T_{d}(Y^{\prime}) = 0.06 + 0.54 = 0.60$, is almost
identical to what was found for the $Y$ point. The effect of disorder
has been to {\em decrease} the transmission for states which were
originally weakly reflected. The strong k-dependence of the
transmission found in the specular case is largely destroyed by a
small amount of disorder in the minority-spin channel. The
contribution from specular component (integrated over 2D BZ) is reduced to 
15\% of the total transmission.

\subsection{Interface resistance}

\begin{figure}[t]
\includegraphics[scale=0.35,clip=true]{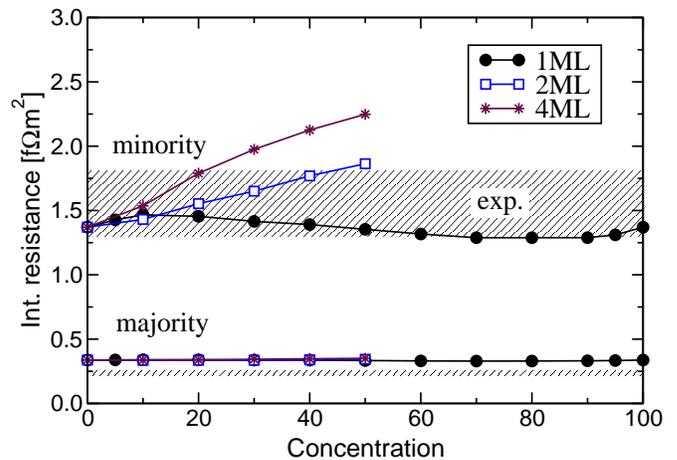}
\caption{Interface resistance for disordered interfaces as a
function of the alloy concentration used to model disordered
interfaces calculated using \eqref{eq:R_Schep} and the
transmission probabilities shown in Fig.~\ref{fig:spec_diff}. The
experimental values for sputtered and MBE grown multilayers cited
in Table I of Ref.~\onlinecite{Bass:jmmm99} span a range of values
which is indicated by the shaded regions.}
\label{fig:int_res}
\end{figure}

To the best of our knowledge, spin-dependent interface transmissions
have not yet been measured directly. What is usually done
\cite{Pratt:prl91,Gijs:prl93} is to measure total resistances for a
whole series of magnetic multilayers in which the total number of
interfaces and/or the thicknesses of the individual layers is varied.
The measured results are interpreted in terms of volume resistivities
and interface resistances. By applying an external magnetic field, the
magnetizations of neighbouring layers which are oriented antiparallel
(AP) can be forced to line up in parallel (P). By measuring the
resistances in both cases, spin-dependent volume resistivities and
interface resistances can be extracted using the two current series
resistor model.\cite{Zhang:jap91,Lee:jmmm93,Valet:prb93} If we take
expression \eqref{eq:R_Schep} which relates the interface transmission
to the interface resistance occurring in the 2CSR model as given,%
\cite{Schep:prb97,Stiles:prb00} we can study how typical uncertainties in interface
transmission, arising from arbitrary assumptions about the interface
disorder, lattice constant or basis set translate into uncertainty in
predicted interface resistances.
Using the transmission probabilities from Fig.~\ref{fig:spec_diff} in
\eqref{eq:R_Schep} results in the curves shown in Fig.~\ref{fig:int_res}.
For comparison, a range of literature values for the spin-dependent
interface resistances derived from experiments on sputtered and MBE
(molecular beam exitaxy) grown multilayers\cite{Bass:jmmm99} is
included in the figure.

\begin{table}[b]
\begin{ruledtabular}
\begin{tabular}{lccc}
\multicolumn{1}{c}{$a (\mathring{A}$)} &
       \multicolumn{2}{c} {3.549} &
                                     \multicolumn{1}{c} {3.614} \\
\hline
\multicolumn{1}{c}{Basis} &
    \multicolumn{1}{c}{\em spdf} &
                      \multicolumn{1}{c}{\em spd} &
                                                 \multicolumn{1}{c}{\em spd} \\
\hline
$R^{\rm maj}(111)$  &  0.46 &  0.39 &  0.34  \\
$R^{\rm min}(111)$  &  1.33 &  1.32 &  1.37
\end{tabular}
\end{ruledtabular}
\caption{Interface resistances, in units of f$ \Omega \mathrm{m}^{2}$,
for ordered interfaces, calculated using expression \eqref{eq:R_Schep}
and the data from Tables \ref{tab:A} and \ref{tab:B}.
The values given here for a lattice constant of $a=3.614\mathring{A}$ differ
slightly from those reported in Ref.~\onlinecite{Xia:prb01} which were
performed using energy-independent muffin-tin orbitals linearized about
the centers of gravity of the occupied conduction states and not at the
Fermi energy. The current implementation\cite{Zwierzycki:prb03} uses
energy-dependent, (non-linearized) MTO's, calculated exactly at the Fermi
energy which improves the accuracy at no additional cost.
}
\label{tab:C}
\end{table}

For the minority-spin case, experimental values (in units of
f$ \Omega \mathrm{m}^{2}$) range from 1.30-1.80 compared to calculated
values of
1.29 for Cu$[$Cu$_{.3}$Co$_{.7}]$Co,
through 1.37 for a disorder-free interface,
to a value of 2.25 for the 4ML model with $x=0.5$,
Cu$[$Cu$_{.83}$Co$_{.17}|$Cu$_{.67}$Co$_{.33}|$Cu$_{.33}$Co$_{.67}|$Cu$_{.17}$Co$_{.83}]$Co.
The influence of lattice constant and basis set on the clean interface
resistance values is small (see Table~\ref{tab:C}).
The present modelling of interface alloying shows that the interface
resistance is more strongly dependent on the detailed spatial
distribution of disorder than was previously found\cite{Xia:prb01}
where only the concentration range $x=0.5 \pm 0.06$ of the 2ML interface
alloy model extracted from experiment%
\cite{deGronckel:prb91,Meny:prb92,Henry:prb96,Kapusta:jac99}
was explored.

\begin{figure}[btp]
\includegraphics[scale=0.35,clip=true]{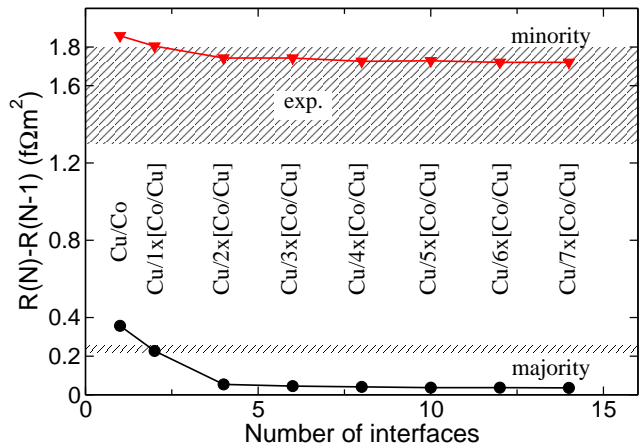}
\caption{Differential interface resistance as the number of
interfaces increase for a disordered Cu/Co(111) multilayer
embedded between Cu leads. A $10 \times 10$ lateral supercell was
used and the interface was modelled as two layers of 50\%-50\%
alloy (2ML model). The results represent an average over 5 disorder configurations
and were obtained for a ``standard'' configuration.\cite{fn:standard}
The range of experimental values\cite{Bass:jmmm99} is indicated by the shaded regions.
}
\label{fig:diff_res}
\end{figure}

For the majority-spin case, the spread in values of the interface
resistance extracted from experiment (for the same samples as for
the minority-spin case) is quite small, 0.22-0.25, and does not
overlap with the values of $0.34$
found for a lattice constant of $a=3.614\mathring{A}$. Unlike the
minority-spin case, changing the lattice constant or using an $spdf$
basis leads to substantially {\em larger} values (Table~\ref{tab:C}).
Because the majority-spin transmission does not depend on the details
of the interface disorder, this cannot be the origin of the discrepancy.
Motivated by the weak scattering in this case, we examine the validity
\cite{Valet:prb93,Gijs:ap97,Bass:jmmm99,Tsymbal:prb00b,Shpiro:prb00}
of the 2CSR model by calculating the resistance of a magnetic multilayer
containing  a large number of disordered interfaces and plot the
resistance added by each additional interface in Fig.~\ref{fig:diff_res}.
Compared to similar calculations in Ref.~\onlinecite{Xia:prb01},
the number of interfaces, size of lateral supercell ($10 \times 10$)
and disorder configurations averaged over are increased substantially.
While the calculations are in very good agreement with Ohm's law for
the strongly scattering minority-spin case, it can be seen that this
is not the case for the majority-spin electrons. For a small number of
interfaces there is a clear breakdown of Ohm's law and thus of the 2CSR
model. The interface resistance eventually saturates at a value much
lower than those extracted from experiment. While inclusion of bulk
scattering will modify this picture somewhat, exploratory calculations%
\cite{Gerritsen:02} indicate that the type of ``bulk'' impurities which
may be reasonably expected to be found in sputtered or MBE grown
multilayers affect the minority spin electrons much more than the
majority spins. Agreement for the latter can only be achieved at the
expense of ruining good agreement for the former.

\section{DISCUSSION}
\label{sec:discussion}

Details of a muffin tin orbital-based method suitable for calculating
from first-principles scattering matrices involving layered magnetic
materials have been given. In a wide range of applications,%
\cite{Xia:prb01,Xia:prb02,Xia:prl02,Zwierzycki:prb03,Bauer:prl04,Zwierzycki:prb05}
it has been shown to be much more efficient and transparent than a
previously used LAPW-based method.%
\cite{vanHoof:97,Schep:prb97,vanHoof:prb99}
Various other schemes have been developed for calculating the
transmission of electrons through an interface (or a more extended
scattering region)
both from first principles,\cite{Stiles:prb88,Stiles:prl91,vanHoof:97,%
MacLaren:prb99,Kudrnovsky:prb00,Riedel:prb01,Taylor:prb01,%
Brandbyge:prb02,Wortmann:prb02a,Wortmann:prb02b,Thygesen:prb03,%
Mavropoulos:prb04,Khomyakov:prb04} or using as input electronic
structures which were calculated from first principles.%
\cite{Tsymbal:jp97,Mathon:prb97a,Mathon:prb97c,Sanvito:prb99,%
Butler:prb01,Mathon:prb01,Mathon:prb05}
Most are based upon a formulation for the conductance in terms of
non-equilibrium Green's functions \cite{Caroli:jpc71a} (NEGF) which
reduces in the appropriate limit to the well known Fisher-Lee (FL)
linear-response form\cite{Fisher:prb81} for the conductance of a finite
disordered wire embedded between crystalline leads. Most implementations
of the NEGF or FL schemes have two disadvantages. (i) The transmission
is calculated for a complex energy which leads to difficulties in
studying for example, tunneling magnetoresistance, where the finite
imaginary part can give rise to an exponential decay which obscures
the interesting physical decay of the transmission as a function of
the barrier thickness. (ii) For a given value of transverse crystal
momentum, the transmission is expressed as a trace over the basis set
in terms of which the Green's function and self-energy are expressed.%
\cite{Khomyakov:prb05} While this has the advantage that the total
transmission can be calculated without explicitly determining the
scattering states and can be computationally efficient, summation of
the contributions from multiple scattering states can obscure real
physical effects, for example, the role of the symmetries of individual
scattering states seen in Fig.~\ref{fig:CuCo111_min}. Explicit
determination of the scattering states not only makes a detailed
analysis of the scattering possible. The full scattering matrix,
expressed in terms of the scattering states, can be used to bridge%
\cite{Xia:prl02} the gap between first-principles electronic structure
calculations and phenomenological models of transport used to analyse
complex situations where a full first-principles treatment is not
practical.

We have instead made use of an alternative technique, suitable for
Hamiltonians that can be represented in tight-binding form, that was
formulated by Ando\cite{Ando:prb91} and is based upon direct matching
of the scattering-region wave function to the Bloch modes of the leads.
The relationship between the wave function matching \cite{Ando:prb91}
and Green function \cite{Caroli:jpc71a,Fisher:prb81} approaches is not
immediately obvious. It was suggested recently that WFM was incomplete
\cite{Krstic:prb02} but the equivalence of the two approaches could be
proven.\cite{Khomyakov:prb05} Schemes similar in spirit to our own, but
based upon empirical tight-binding Hamiltonians have been presented by
Sanvito {\em et al.}\cite{Sanvito:prb99} and by Velev.%
\cite{Velev:prb03,Velev:prb04} In contrast to these schemes, our TB-MTO
formalism is a parameter-free approach that has all of the advantages
derived from self-consistent determination of potentials and spin
densities for systems for which these are not known from experiment.
Judging from the size of systems to which it has been applied, it would
seem that our implementation is nevertheless substantially more efficient
than these empirical schemes. The scattering regions treated in
Figs.~\ref{fig:config_av},\ref{fig:spec_diff},\ref{fig:dis_scatt_maj} and
\ref{fig:dis_scatt_min} contained as many as 3200 atoms
($20 \times 20$ lateral supercell $\times $ 8 principal layers where the
potential was allowed to deviate from its bulk values) or, in the case of
Fig.~\ref{fig:diff_res},
$\sim 15000$ atoms ($10 \times 10$ lateral supercell $\times$ 150
principal layers).
Our WFM scheme should not be confused\cite{Velev:prb04} with a recently
developed transport formalim\cite{Kudrnovsky:prb00,Drchal:prb02} also
based upon TB-LMTOs but which makes use of the Caroli NEFG expression
for the conductance in terms of a trace and a complex energy. Khomyakov
and Brocks\cite{Khomyakov:prb04} have developed a scheme analogous to
ours but based upon pseudopotentials and a real space grid which make
it more suitable for studying quantum wires or the type of open
structures studied in molecular electronics, but is computationally
much more expensive.

A third approach based upon
``embedding''\cite{Inglesfield:jpc81,Crampin:jp92} has been combined
with full-potential linearized augmented plane wave method to
yield what is probably the most accurate scheme to date
\cite{vanHoof:97,Wortmann:prb02a,Wortmann:prb02b} but like the real
space grid WFM method,\cite{Khomyakov:prb04} these methods are
numerically very demanding.


\section{SUMMARY}
\label{sec:summary}

Details of a wave-function matching method suitable for calculating the
scattering matrices in magnetic metallic hybrid structures based upon
first-principles tight-binding muffin tin orbitals have been given and
illustrated with calculations for a variety of Co/Cu(111) interface-related
problems. The minimal basis of localized orbitals is very efficient, allowing
large lateral supercells to be handled. This allow us to model materials with
large lattice mismatch or to study transport in the diffusive regime. Because
the scattering states are calculated explicitly, the effect of various types
of scattering can be analyzed in detail.

\begin{acknowledgments}
This work is part of the research program for the
``Stichting voor Fundamenteel Onderzoek der Materie'' (FOM)
and the use of supercomputer facilities was sponsored by the
``Stichting Nationale Computer Faciliteiten'' (NCF), both financially
supported by the
``Nederlandse Organisatie voor Wetenschappelijk Onderzoek'' (NWO).
It was also supported by the European Commission's
Research Training Network {\em Computational Magnetoelectronics}
(contract No. HPRN-CT-2000-00143) as well as by the
NEDO International Joint Research program {\em Nano-scale Magnetoelectronics}.
MZ wishes also to acknowledge support from KBN grant
No.~PBZ-KBN-044/P03-2001.
We are grateful to: Ilya Turek for his TB-LMTO-SGF layer code which we
used to generate self-consistent potentials and for numerous discussions
about the method; Anton Starikov for permission to use his version of the
TB-MTO code based upon sparse matrix techniques to perform some of the
calculations.
\end{acknowledgments}

\appendix

\section{Velocities}
\label{sec:velocities}

Expressions for the velocities of the propagating modes in the
leads (Sect.~\ref{ssec:Leads}) are more easily derived using an
energy independent Hamiltonian than the energy dependent
tail-cancellation condition of section \ref{ssec:MTOs}. To do so,
we make use of the close relationship between the KKR
tail-cancellation equation (\ref{eq:TBPmS}) and the linearized MTO
(LMTO) Hamiltonian, both of which can be expressed in terms of the
Hermitian matrix\cite{Andersen:85,Andersen:prb86}
\begin{gather}
  h^\alpha(\varepsilon)
  =- [\dot{P}^\alpha(\varepsilon)]^{-1/2}
      \left(P^\alpha(\varepsilon)-S^\alpha \right)
     [\dot{P}^\alpha(\varepsilon)]^{-1/2}          \nonumber \\
  = -P^\alpha(\varepsilon) [\dot{P}^\alpha(\varepsilon)]^{-1}
  + [\dot{P}^\alpha(\varepsilon)]^{-1/2} S^\alpha
               [\dot{P}^\alpha(\varepsilon)]^{-1/2}.
  \label{eq:halpha}
\end{gather}
Fixing the energy at $\varepsilon=\varepsilon_F$
and defining the potential parameters\cite{Andersen:prb86,Andersen:87}
\begin{subequations}
\label{eq:potpar}
\begin{equation}
\sqrt{d^\alpha}=[\dot{P}^\alpha(\varepsilon_F)]^{-1/2}
\label{subeq:1}
\end{equation}
\begin{equation}
c^\alpha=-P^\alpha(\varepsilon_F)/ \dot{P}^\alpha(\varepsilon_F) + \varepsilon_F,
\end{equation}
\end{subequations}
(\ref{eq:halpha}) can be written as
\begin{equation}
 h^\alpha \equiv h^\alpha(\varepsilon_F) = c^\alpha
          + \sqrt{d^\alpha} \; S^\alpha \, \sqrt{d^\alpha}-\varepsilon_F
  \label{eq:tbham}
\end{equation}
Equation (\ref{eq:tbham}) has the form of a two-center tight binding
Hamiltonian whose energy is given relative to $\varepsilon_F$. It
provides the lowest order approximation\cite{Andersen:85,Andersen:prb86}
to the full LMTO Hamiltonian and yields eigenvalues correct to first order
in $(\varepsilon-\varepsilon_F)$.
For eigenvalues equal to $\varepsilon_F$, it yields eigenvectors which
are equal to those determined by the tail-cancellation condition
(\ref{eq:TBPmS}), up to a scaling factor $(\dot{P}^\alpha)^{-1/2}$.

To calculate the group velocities of states precisely at the
linearization energy, in the present case at the Fermi energy,
the first-order Hamiltonian (\ref{eq:tbham}) can be used since any
error vanishes identically for $\varepsilon(\mathbf{k})=\varepsilon_F$.
Using the translational symmetry of the leads, the Hamiltonian
(\ref{eq:tbham}) for Bloch vector ${\bf k}$ is
\begin{equation}
  h^\alpha_{RL,R'L'}(\mathbf{k})=
  \sum_{\mathbf{T}}e^{i\mathbf{k}\cdot\mathbf{T}}
  h^\alpha_{RL,(R'+T)L'}
  \label{eq:tbham_bloch}
\end{equation}
where $RL$ labels the sites and orbitals within the unit cell
and $\mathbf{T}$ runs over lattice vectors.
The energy eigenvalues $\varepsilon_{\mu}(\mathbf{k})$ are the
expectation values
\begin{equation}
  \varepsilon_{\mu}(\mathbf{k})=\mathbf{a}^\dagger_{\mu}(\mathbf{k})
  h^{\alpha}(\mathbf{k}) \mathbf{a}_{\mu}(\mathbf{k})
  \label{eq:expect_val}
\end{equation}
where the eigenvectors $\mathbf{a}_{\mu}(\mathbf{k})$ are indexed by $RL$ and
we assumed normalization $\mathbf{a}^\dagger_{\mu}\cdot\mathbf{a}_{\mu}=1$.
It is now straightforward to calculate the group velocity of the propagating
mode
\begin{equation}
  \begin{split}
 {{\bm \upsilon }}_{\mu}=\frac{1}{\hbar}
  \frac{\partial \varepsilon_{\mu}(\mathbf{k})}{\partial \mathbf{k}}
  & = \frac{i}{\hbar}\sum_{\mathbf{T}}
                    \mathbf{T}e^{i \mathbf{k} \cdot \mathbf{T}} \times  \\
  & \sum_{RL,R'L'}  a^{*}_{RL}  h^{\alpha}_{RL,(R'+T)L'} a_{R'L'} \\
  \end{split}
  \label{eq:3dvelocity}
\end{equation}
In the mixed representation $\left|I,\mathbf{k}_{\parallel}\right>$ defined
in section \ref{ssec:Leads} \eqref{eq:3dvelocity} gives for the
velocity in the stacking direction
\begin{equation}
  \upsilon_{\mu}=\frac{id}{\hbar}\left[
    \mathbf{a}^\dagger_{\mu} h^\alpha_{I,I+1}(\mathbf{k}_{\parallel})
    \lambda_{\mu} \mathbf{a}_{\mu}-\mathbf{h.c.}
    \right]
  \label{eq:layered_velociy}
\end{equation}
where $d$ is the distance between equivalent monolayers in adjacent
principal layers (PL), the hopping is assumed (as in section
\ref{ssec:Leads}) to extend only between neighbouring PLs and
$\lambda_{\mu}=exp(i\mathbf{k}\cdot\mathbf{T}^0)$ with $\mathbf{T}^0$
connecting equivalent sites in the neighbouring PLs.
Using the definition \eqref{eq:halpha} of $h^\alpha$  and
recalling that the solutions $\mathbf{u}_{\mu}$ of the
tail-cancellation equation \eqref{eq:TBPmS} take implicitly into
account the scaling factor $(\dot{P}^\alpha)^{-1/2}$ we arrive at
equation \eqref{eq:velocity}.

\section{Symmetry relations}
\label{sec:symmetry}

If we look closely at the transmission probabilities in
Fig.~\ref{fig:CuCo111_min}, we see that the sheet resolved
transmissions exhibit the geometrical symmetry of the underlying
lattice (\emph{i.e.} the three-fold rotational axis). The total
transmission probability on the other hand possesses an extra
inversion symmetry,
$T({\bf k_{\parallel }})=T(-{\bf k_{\parallel }})$, which results in
plots with a six-fold rotational axis. This higher symmetry is the
manifestation of the fundamental time-reversal symmetry obeyed in
the absence of spin-orbit coupling and a magnetic field. In
the case of the bulk system time-reversal symmetry grants that for
every eigenstate $\psi_\alpha({\bf k})$ there exists the
counterpart with the same energy and opposite wave vector
(\emph{i.e.} $\varepsilon_\alpha({\bf k})=\varepsilon_\alpha(-{\bf
k})$) and the wave functions are related by the complex conjugate.
The situation is more complicated in the case of the scattering
state. Consider a state incoming from the
left lead and  scattered in the middle region. The wave function
consists then of the incoming and reflected states in the left
lead
\begin{equation}
  \Psi^r_{\mathcal{L}}({\bf k_{\parallel }})=\psi^{+}_{\mu}({\bf k_{\parallel }})+
  \sum_{\mu'}r_{\mu'\mu}({\bf k_{\parallel }})\psi^{-}_{\mu'}({\bf k_{\parallel }})
  \label{eq:left_ret}
\end{equation}
and of the transmitted states in the right lead
\begin{equation}
  \Psi^{r}_{\mathcal{R}}({\bf k_{\parallel }})
  =\sum_{\nu}t_{\nu\mu}({\bf k_{\parallel }})\psi^+_{\nu}({\bf k_{\parallel }}).
  \label{eq:right_ret}
\end{equation}
The time reversal operation transforms the above ``retarded'' state into the
``advanced'' one in which a number of  incoming states (from the left
and the right) combine to produce a single outgoing state on the
left, \emph{i.e.}
\begin{equation}
  \Psi^a_{\mathcal{L}}(-{\bf k_{\parallel }})=
  \sum_{\mu'}r^{*}_{\mu'\mu}({\bf k_{\parallel }})\psi^{+}_{\mu'}(-{\bf k_{\parallel }})
  +\psi^{-}_{\mu}(-{\bf k_{\parallel }})
  \label{eq:left_adv}
\end{equation}
and
\begin{equation}
  \Psi^{a}_{\mathcal{R}}(-{\bf k_{\parallel }})=
  \sum_{\nu}t^{*}_{\nu\mu}({\bf k_{\parallel }})\psi^-_{\nu}(-{\bf k_{\parallel }}).
  \label{eq:right_adv}
\end{equation}
Equations (\ref{eq:left_adv}) and (\ref{eq:right_adv}) impose a set
of conditions on the values of scattering coefficients for the states
with $-{\bf k_{\parallel }}$. Combined with the analogous conditions derived
for the states with the incoming state in the right lead, they are
compactly expressed as
\begin{equation}
  I=S\left(-{\bf k_{\parallel }}\right)S^{*}\left({\bf k_{\parallel }}\right)\;\;\Rightarrow\;\;
  S\left(-{\bf k_{\parallel }}\right)=S^{T}\left({\bf k_{\parallel }}\right).
\label{eq:scatt_symm}
\end{equation}
The scattering matrix $S$ is defined as
\begin{equation}
  S=\left(
    \begin{array}{cc}
      r & t' \\
      t & r'
    \end{array}
  \right)
  \label{eq:scatt_def}
\end{equation}
where $r^{(}{'}^{)}$ and  $t^({'}^)$ are matrices in the space of the lead
modes and the primed coefficients describe scattering of the states
incoming from the right. More specifically we have:
\begin{equation}
  t_{\nu\mu}(-{\bf k_{\parallel }})=t'_{\mu\nu}({\bf k_{\parallel }})
  \;\;\mathrm{and}\;\;
  r_{\mu'\mu}(-{\bf k_{\parallel }})=r_{\mu\mu'}({\bf k_{\parallel }})
  \label{eq:tr_symm}
\end{equation}
Equation~\eqref{eq:tr_symm} gives
\begin{equation}
  T_{\mathcal{LR}}(-{\bf k_{\parallel }})
 =\sum_{\nu\mu} |t_{\nu\mu}(-{\bf k_{\parallel }})|^2
 =\sum_{\mu\nu}|t'_{\mu\nu}({\bf k_{\parallel }})|^2
 =T_{\mathcal{RL}}({\bf k_{\parallel }})
\end{equation}
In addition, for any two-terminal device, the Hermiticity of the
scattering matrix guarantees that
$T_{\mathcal{RL}}({\bf k_{\parallel }})=T_{\mathcal{LR}}({\bf k_{\parallel}})$
(see Ref.~\onlinecite{Datta:95}) which finally proves the in-plane
inversion symmetry mentioned at the beginning. The last step can
not however be taken for the partial (FS resolved) transmission
probabilities. These quantities thus possess only the geometrical
symmetry of the system.


\begin{thebibliography}{85}
\bibitem[{fn:({\natexlab{a}})}]{fn:total_en}
\bibinfo{note}{Because the magnetism of transition metals depends very
  sensitively on atomic structure,\cite{Kubler:00} it is important to know this
  structure quite accurately. The current drive to make devices whose lateral
  dimensions approach the nanoscale means that it is becoming increasingly
  important to know the atomic structures of these small systems
  microscopically while at the same time it is more difficult to do this
  characterization experimentally. It has become a practical alternative to
  determine minimum-energy structures theoretically by minimizing as a function
  of the atomic positions the total energy obtained by solving the
  Schr{\"{o}}dinger equation self-consistently within the local density
  approximation (LDA) of Density Functional Theory (DFT), thereby avoiding the
  use of any free parameters}.
\end{thebibliography}
\end{document}